\documentclass[12pt,epsf]{iopart}
\usepackage{iopams}
\usepackage{graphics}
\usepackage{graphicx}
\newcommand{\text}{\hbox}

\newcommand{\eps}{\varepsilon}
\newcommand{\p}{\partial} 
\newcommand{\dd}{\hbox{d}}
\newcommand{\ee}{\hbox{e}}

\newcommand{\rD}{\mathcal{D}}

\newcommand{\nav}{\bar{n}} 
\newcommand{\vn}{\vec{n}}

\newcommand{\hH}{\hat{H}}

\newcommand{\begeq}{\begin{equation}}
\newcommand{\eqend}{\end{equation}}
\def\lsim{\:\raisebox{-0.5ex}{$\stackrel{\textstyle<}{\sim}$}\:}

\begin{document}

\title[]{Segregation in diffusion-limited multispecies pair annihilation}

\author{H.J. Hilhorst\dag, O. Deloubri\`ere\ddag, M.J. Washenberger\ddag, 
	and U.C. T\"auber\ddag}

\address{\dag\ Laboratoire de Physique Th\'eorique, B\^atiment 210
Universit\'e de Paris-Sud, F-91405 Orsay cedex, France}

\address{\ddag\ Department of Physics, Virginia Polytechnic Institute and
State University, Blacksburg, Virginia, 24060, USA}

\begin{abstract} 
The kinetics of the $q$ species pair annihilation reaction ($A_i+A_j\to
\varnothing$ for $1\leq i < j \leq q$) in $d$ dimensions is studied by means
of analytical considerations and Monte Carlo simulations. In the long-time
regime the total particle density decays as $\rho(t)\sim t^{-\alpha}$. For 
$d=1$ the system segregates into single species domains, yielding a different 
value of $\alpha$ for each $q$; for a simplified version of the model in one
dimension we derive $\alpha(q)=(q-1)/(2q)$. Within mean-field theory, 
applicable in $d\geq 2$, segregation occurs only for $q<1+(4/d)$. The only 
physical realisation of this scenario is the two-species process ($q=2$) in 
$d=2$ and $d=3$, governed by an extra local conservation law. For $d\geq 2$ and
$q\geq 1+(4/d)$ the system remains disordered and its density is shown to decay
universally with the mean-field power law ($\alpha=1$) that also characterises
the single-species annihilation process $A+A\to \varnothing$.
\end{abstract}




\section{Introduction}
\subsection{Motivation}

Thermal equilibrium is a very special situation and nature provides us with 
numerous examples of systems that cannot be described by equilibrium 
statistical thermodynamics. Investigation of simple models is necessary to 
understand the various dynamical phenomena observed in out-of-equilibrium 
systems, such as collective behaviour, phase transitions, or self-ordering 
\cite{chopard98,marro99}. Among these models reaction-diffusion processes play 
a paradigmatic role because of their simple definition on the microscopic level
and the complex phenomena they may exhibit on the macroscopic level
\cite{smoluchowski16,ovchinnikov78,toussaint83}. Reaction-diffusion processes 
may describe not only chemical reactions \cite{kuzovkov88,ovchinnikov89}, but 
also {\it e.g.} phase transitions in ionic conductors \cite{dieterich80}, 
epidemic spreading \cite{molisson77}, or forest fires \cite{albano94}. The 
study of such simply defined models is justified by the fact that they entail
macroscopic phenomena that are {\it universal,} {\it i.e.} that in the regime 
of asymptotically long times and large distances do not depend on the details 
of microscopic dynamics (for recent reviews, see 
Refs.~\cite{hinrichsen00,odor03}).

\subsection{Annihilation models}
\label{ann}

The simplest nontrivial diffusion-limited reaction is the {\it single-species
pair annihilation process} where random walkers (particles of type $A$) 
annihilate each other according to $A+A\to \varnothing$ when they meet. Here 
$\varnothing$ stands for an inert species that plays no further role in the 
dynamics. The diffusion-limited coagulation process $A+A\to A$ displays the
same asymptotic behaviour \cite{peliti86}. An experimental realisation is the 
diffusion and annihilation of photo-generated excitons on tetramethylammonium 
manganese trichloride \cite{kroon93}. In the long-time regime, for systems 
with spatial dimension $d$, the particle density decays as $\rho(t)\sim t^{-1}$
for $d>2$, which is the mean-field result, whereas $\rho(t)\sim t^{-d/2}$ for 
$d<2$. Exactly at the upper critical dimension $d_c=2$ one finds logarithmic 
corrections $\rho(t)\sim t^{-1} \ln t$. 

These asymptotic power laws were first obtained by means of exact calculations 
for specific models \cite{bramson80}. Their universality is established through
a mapping of the corresponding classical master equations onto a stochastic 
field theory \cite{doi76,grassberger80,peliti85}, and its subsequent analysis 
by means of the dynamical renormalisation group (RG) \cite{peliti86,lee94}. In 
the field theory language, the simple annihilation and coagulation processes 
are characterised by the absence of any propagator renormalisation, whence the
relevant length scale is the diffusion length $\ell(t) \sim t^{1/2}$. The 
reaction rate enters as a nonlinear vertex, whose perturbative renormalisation 
can be summed to all orders by means of a Bethe-Salpeter equation. At long 
length and time scales, the renormalised reaction rate approaches a universal 
RG fixed point for $d<2$. The solution of the Callan-Symanzik equation for the 
particle density then yields the aforementioned results \cite{lee94}. 

The {\it two-species annihilation process} $A+B\to \varnothing$ (with no 
reactions occurring between alike particle types) exhibits interesting new 
phenomena, especially for $d<4$ \cite{ovchinnikov78,toussaint83,lee95}. For a 
random initial distribution of $A$ and $B$ particles with equal densities
$\rho_A(0)=\rho_B(0)$, in the course of time segregation into monospecies 
domains emerges. The average domain size is given by the typical diffusion 
length, and hence grows as $\ell(t)\sim t^{1/2}$. The reactions happen only in 
narrow zones between these domains. As a consequence the total particle density
decays more slowly than in the single-species process: Whereas the homogeneous 
mean-field rate equation solution $\rho(t) \sim t^{-1}$ holds for $d\geq 4$, 
one has $\rho(t)\sim t^{-d/4}$ for $d<4$. The reaction zone width 
$\ell_{\rm int}(t)$ also tends to infinity according to a power law, 
$\ell_{\rm int}(t)\sim t^{\lambda_r}$, with, in particular, 
$\lambda_r=3/8$ in $d=1$
\cite{leyvraz92}. The renormalisation group analysis for this two-species 
process establishes that the upper critical dimension in the RG sense is still 
$d_c=2$, {\it i.e.}, mean-field rate equations for the local particle densities
augmented with diffusion to account for spatial variations provide an apt 
description for $d>2$ \cite{lee95}. Thus, the borderline dimension for 
segregation $d_{\rm seg}=4$ is located inside the realm of validity of 
mean-field theory. 

For unequal initial densities of the $A$ and $B$ species 
\cite{kang84,bramson91}, the final state will be characterised by a nonzero 
density of the majority species. Above $d_c=2$ the density 
$\rho_{\text{{\tiny min}}}$ of the minority species decays exponentially. For 
$d\leq d_c$ this decay becomes a stretched exponential with 
$\ln \rho_{\text{{\tiny min}}}\sim -{t}/{\ln t}$ for $d=2$ and
$\ln \rho_{\text{{\tiny min}}}\sim -t^{d/2}$ for $d<2$.

It is worth mentioning that certain special initial conditions with long-range 
correlations are capable of inhibiting segregation. As an example, consider an
alternating alignment of particles $\ldots ABABAB\ldots$ in one dimension, and 
reactions with infinite rate. Then any encounter will involve particles of 
distinct species, leading to their annihilation. Yet removing any neighbouring 
$AB$ pair will conserve the alternating alignment. Since alike particles can 
therefore never meet, this process is in fact equivalent to the single-species 
reaction $A+A\to \varnothing$. Segregation is impossible, and the total density
decays as $\sim t^{-1/2}$ \cite{krapivsky00}.

It is now natural to view the two above models as special cases of a 
Multispecies Annihilation Model (MAM) \cite{benavraham86,deloubriere02,zhong03}
in which $q$ different species $A_1,A_2,\ldots,A_q$ perform random walks on a 
lattice and, when meeting on the same lattice site, may react according to
\begin{equation}
A_i+A_j\to \varnothing \ , \quad i\neq j \, .
\end{equation}
We will consider that the hopping rates and the reaction rates are 
species-independent ({\it i.e.}, we assume uniform diffusion constants $D_i=D$ 
and reaction rates $\lambda_{ij}=\lambda$). For equal initial densities 
$\rho_i(0)=\rho(0)/q$ one expects a power law decay
\begin{equation}
\rho(t)\sim t^{-\alpha} ,
\label{defalpha}
\end{equation} 
with a $q$- and $d$-dependent exponent that we denote as $\alpha(q,d)$, or,
when we are interested only in its $q$ dependence, as $\alpha(q)$. In this 
paper we shall investigate under which conditions the MAM may exhibit species
segregation, and we will determine the value of the exponent $\alpha$ as 
function of $q$ and $d$. Thus we fully characterise the long-time behaviour of 
a multicomponent reaction-diffusion system for arbitrary number of species, and
elucidate the physics behind the different asymptotic scaling laws. Similar to
other multispecies systems with hard-core constraints \cite{odor02}, the MAM 
process turns out to display very special features in one dimension, owing to 
the special topological limitations of hopping on chains.

One special limit of the MAM may be analysed immediately. For an infinite 
number of species, $q=\infty$, the probability for two particles of the same 
type to meet is zero. Therefore all encounters lead with rate $\lambda$ to a 
reaction. Distinguishing different particle species then loses its meaning, 
whence for $q=\infty$ we recover again the single-species pair process 
$A+A\to \varnothing$ in $d$ dimensions \cite{benavraham86}.

We remark that novel behaviour can be expected generically only in the highly
symmetric case with uniform reaction rates and equal initial densities. In any
other situation, one would expect that at long times only the least reactive
and/or most populous species will survive. Once the minority particle species
have disappeared, one should expect the process to be described by the 
$A+B\to \varnothing$ reaction with unequal particle densities 
\cite{deloubriere02}.

\subsection{Results}

We shall present numerical and analytical results concerning the asymptotic 
decay of the total particle density $\rho(t)$ of the $q$-species MAM in $d$
dimensions. There are two physical effects that invalidate the homogeneous 
mean-field rate equation predictions in low dimensions, namely reaction rate 
renormalisation and potentially species segregation. Associated with these are 
two borderline dimensions. First, below the {\it upper critical dimension} 
$d_c=2$ (independent of $q$) fluctuations in the dynamics are relevant in the 
RG sense, renormalising the effective reaction rate $\lambda$. Second, within 
the mean-field approximation, below a $q$-dependent {\it segregation dimension}
$d_{\rm seg}$ given by 
\begin{equation}
d_{\rm seg}(q) = \frac{4}{q-1} \ ,
\label{resultdseg}
\end{equation}
the system splits up into monospecies domains. The behaviour predicted by the 
homogeneous mean-field rate equations (that do not allow for segregation) is 
observed only for $d>\max(d_c,d_{\rm seg})$. 

The full picture of how the $q$-species MAM behaves in $d$ dimensions then 
emerges as follows:
\begin{itemize}

\item For $d=1$ the special topological constraints of a linear chain produce
segregation with a nonuniversal value of the decay exponent that we determine
\cite{deloubriere02,zhong03} to be
\begin{equation}
\alpha(q,1)=\frac{q-1}{2q} \ , \qquad d=1 \, .
\label{eqnalphaq1}
\end{equation}
This expression of course reproduces the known one-dimensional exponents 
$\alpha(2,1)=1/4$ and $\alpha(\infty,1)=1/2$. 

\item For $d\geq 2$ and $q=2$ an extra conservation law in the microscopic 
dynamics causes segregation and 
\begin{equation}
\alpha(2,d)=\frac{d}{4}\ , \qquad 2\leq d\leq d_{\rm seg}(2)=4 \, .
\label{eqnalpha2d}
\end{equation}
Note that when $d=1$ is substituted in Eq.~(\ref{eqnalpha2d}), the result  
coincides with $\alpha(2,1)$ as given by Eq.~(\ref{eqnalphaq1}). The local
conservation law for the particle density difference in two-species 
annihilation $A+B\to \varnothing$ \cite{toussaint83} renders the case $q=2$ 
quite special.

\item If one is willing to analytically continue the results to noninteger $q$,
then Eq.~(\ref{eqnalpha2d}) is a special case of
\begin{equation}
\alpha(q,d)=\frac{(q-1)d}{4}\ , \qquad 2\leq d\leq d_{\rm seg}(q), \quad
1<q<3 \, .
\label{eqnalphaqnoninteger}
\end{equation}
For all $(q,d)$ in this range there occurs segregation, although for $q\neq 2$ 
it cannot be traced back to a conservation law.

\item For $d\geq 2$ and $q=3,4,\ldots,\infty$ there is no segregation. The 
density decay therefore follows a universal (mean-field) power law, independent
of the number of species: 
\begin{equation}
\alpha(q,d)=1, \qquad d\geq 2 , \quad q\geq 3 \, .
\label{eqnalphaqd}
\end{equation}
In the borderline case $d=2$ the power law defined by Eq.~(\ref{eqnalphaqd}) 
is accompanied by a logarithmic correction, $\rho(t)\sim t^{-1} \ln t$, 
precisely as for $A+A\to \varnothing$.
\end{itemize}

In one dimension, when segregation occurs, domains of identical particles are 
separated by {\it reaction zones} where the annihilation reactions takes place.
Let $\ell_{\rm int}$ be the typical distance between closest-neighbour pairs 
$A_i A_j$ in a reaction zone. We will now assume an infinite reaction rate, so 
that there is one such pair per zone and $\ell_{\rm int}$ is also the zone 
width. The temporal growth of $\ell_{\rm int}$ may be analysed as follows. The 
density of zones is equal to $1/\ell(t)$. Hence we may write the rate of 
decrease of the total density as 
\begin{equation}
\frac{\dd\rho(t)}{\dd t} = -\frac{\kappa(t)}{\ell(t)} ,
\label{drhodt1}
\end{equation}
where $\kappa(t)$ is the typical number of annihilations per unit time in a
reaction zone. This relation, which may be taken as the definition of 
$\kappa(t)$, was proposed and exploited by Leyvraz and Redner \cite{leyvraz92}.
The asymptotic time dependence of $\kappa(t)$ is connected by
Eq.~(\ref{drhodt1}) to those of $\rho(t)$ and $\ell(t)$. The time needed for 
the pair $A_i A_j$ to annihilate is $\sim \ell_{\rm int}^2$, so the change of 
particle density per unit time is 
$\dd\rho/\dd t\sim 1/(\ell\, \ell_{\rm int}^2)$. If we now set 
$\rho(t)\sim t^{-\alpha}$ and $\ell_{\rm int}\sim t^{\lambda_r}$, with 
$\ell\sim t^{1/2}$, then Eq.~(\ref{drhodt1}) yields the scaling relation 
\cite{deloubriere02,zhong03}
\begin{equation}
\lambda_r(q) = \frac12 \left[ \alpha(q)+\frac12 \right] = \frac{2q-1}{4q} \ , 
\qquad d=1 \, .
\label{expsca}
\end{equation} 
For the two-species model one recovers $\ell_{\rm int}\sim t^{3/8}$ 
\cite{leyvraz92}.

This paper is organised as follows. In Sec.~\ref{secRG} we recall the known 
methods and results for the $q=2$ and $q=\infty$ cases in a field-theoretic 
perspective. We then extend these arguments to any value of $q$. We show why 
it is difficult in $d<2$ to obtain results for general $q$ by renormalisation 
methods. In Sec.~\ref{secSMAM} we change our approach and consider dimension 
$d=1$ specifically. We introduce a simplified version of the MAM, to be 
referred to as SMAM, in which the stochastic dynamics is replaced by a 
deterministic algorithm. This SMAM is exactly solvable and produces the 
asymptotic density exponent (\ref{eqnalphaq1}). We argue that this result is 
valid also for the original stochastic MAM. Our analytical findings are 
supported by our own Monte Carlo simulations, presented  in Sec.~\ref{secMC}, 
as well as by recent data by Zhong {\it et al.} \cite{zhong03}. We show that 
great caution must be exercised when measuring decay exponents, since crossover
regimes can be long (and depend on the ratio of the reaction rate and the 
diffusion constant). Sec.~\ref{secconclusion} contains some concluding remarks.
Appendix A details the solution of the mean-field rate equations. In Appendix B
we point out the topological specificities of the one-dimensional case by 
mapping the model onto an anisotropic antiferromagnetic Heisenberg spin chain.

\section{Analytical considerations}
\label{secRG}

\subsection{Mean-field theory}
\label{subsecmeanfield}

The kinetics of chemical reactions may in first approximation be investigated 
by mean-field theory. The mean-field rate equations neglect any spatial
variations of the particle densities and thus cannot capture reaction-induced
noise and diffusion-induced correlations. In low dimensions, however, these 
effects may govern the long-time behaviour of the reaction kinetics, 
invalidating the mean-field approximation.

For the MAM, the mean-field rate equations (with a uniform reaction rate
absorbed into the time scale) read 
\begin{equation} 
\frac{\dd \rho_i(t)}{\dd t} = -\sum_{j=1 \atop j\neq i}^{q}
 \rho_i \rho_j, \quad i=1,2,\ldots,q \, .
\label{rateequations}
\end{equation}
For equal initial densities $\rho_i(0)=\rho^*$, Eqs.~(\ref{rateequations}) 
are easily solved to give $\rho(t)=rho^*/[1+(q-1)\rho^*t]$.
Above some critical dimension, $d > \max(d_c,d_{\rm seg})$ therefore expect the
MAM decay exponent $\alpha$ to take on the mean-field value $\alpha(q,d)=1$.

It is of interest to consider Eqs.~(\ref{rateequations}) when the permutation 
symmetry of the species is broken by the initial conditions, {\it i.e.}, when 
the $\rho_i(0)$ are not all equal. For $k=1,\ldots,p$, let there be initial 
values $\rho_k(0) = \rho_k^*$ having multiplicity $n_k$, where
$\sum_{k=1}^p n_k = q$. We then arrive at a set of $p>1$ independent mean-field
equations which, after renumbering the species and absorbing the rate $\lambda$
into the time scale, may be written as
\begin{equation}
\frac{\dd\rho_k(t)}{\dd t} = -\rho_k 
\left( \sum_{\ell=1}^p n_\ell\rho_\ell-\rho_k \right) , \quad k=1,\ldots,p\, , 
\label{mfnonsymmeq}
\end{equation}
with the initial conditions now all distinct and ordered such that
\begin{equation}
0<\rho_1^*<\rho_2^*<\ldots<\rho_p^* \, .
\label{orderincond}
\end{equation}
The study of this nonlinear system of equations was initiated by Ben-Avraham 
and Redner \cite{benavraham86}.
In Appendix A we complete their analysis and
determine the full analytic solution of the equations 
(\ref{mfnonsymmeq}) and (\ref{orderincond})
for general $p$ and $\{n_\ell\}$. 
The asymptotic large time behaviour may be summarised as follows. 
\begin{itemize}
\item For $n_p>1$ the density $\rho_p$ of the most abundant
species is $n_p$-fold degenerate. 
It decays to zero as $t^{-1}$. The densities of the other species
tend to zero with a faster power common to all of them, {\it viz.} as 
$t^{-1-1/(n_p-1)}$.
\item For $n_p=1$ the density of the most abundant species, $\rho_p$, is 
nondegenerate. It tends exponentially to a constant $\rho_p^*A^*$ with $A^*$ 
given by
\begin{equation}
A^*=\prod_{\ell=1}^{p-1}(1-\rho_\ell^*/\rho_p^*)^{n_\ell} .
\label{defA0}
\end{equation}
The densities of all other species tend exponentially to zero for 
asymptotically large times on 
the common time scale $\tau=1/\rho_p^*A^*$.
\end{itemize}

\subsection{A Ginzburg--type criterion for the segregation dimension 
            $d_{\rm seg}$}
\label{subsecGinzburg}

The homogeneous mean-field approximation of the preceding subsection 
{\it \ref{subsecmeanfield}} is based on the hypothesis of there being no 
particle species segregation. Still within the framework of mean-field theory, 
but allowing now for particle density inhomogeneities, we will provide a 
criterion for the appearance of segregation. To this end, we investigate local 
deviations from the global density averages. When it will appear that these 
deviations are of the same order as the averages, one must conclude that the 
no-segregation hypothesis becomes invalidated. 

Let there be an infinite system with equal initial densities $\rho^*$ for all 
$q$ species. We will consider a finite subvolume $L^d$ of this system. Within 
such a volume the initial densities, due to random fluctuations, will be
\begin{equation}
\rho_{iL}(0)=\rho^* + c_i \sqrt{\rho^*} \, L^{-d/2} \, , \quad 
i=1,2,\ldots,q \, ,
\label{rhoiL0}
\end{equation}   
where the $c_i$ are random constants of order unity with zero average. For 
times $t$ small compared to the diffusion time across the volume, 
$t\lsim t_L\sim L^2/D$, this volume may be considered as isolated from the rest
of the system. We now assume the mean-field approximation to be valid. 
Generically all $c_i$ will be different. The $\rho_{iL}(t)$ then evolve in time
according to Eq.~(\ref{mfnonsymmeq}) with all $n_k=1$ and initial conditions 
given by Eq.~(\ref{rhoiL0}). Let $c_q$ be the largest one of the random 
constants. Hence $\rho_{iL}(t)$ will tend to zero for all $i=1,\ldots,q-1$, 
whereas $\rho_{qL}(t)$ will approach the limit 
$\rho_{qL}(\infty)=\rho^*A_L^*$, with, according to Eq.~(\ref{defA0}),
\begin{eqnarray}
A_L^*&=&\prod_{i=1}^{q-1}[1-\rho_{iL}(0)/\rho_{qL}(0)]\nonumber\\&\simeq&
\left[ \prod_{i=1}^{q-1}(c_q-c_i) \right] (\rho^*)^{-(q-1)/2} \, L^{-(q-1)d/2}
\simeq C \, L^{-(q-1)d/2} \, .
\label{A0L}
\end{eqnarray}
Therefore in an isolated volume $L^d$ only a single species will be left over, 
with density $\sim L^{-(q-1)d/2}$. The time scale for the approach of this 
limiting value is $\tau_L=1/\rho^*A_L^* \sim L^{(q-1)d/2}$. This time scale
is well within the time that the volume stays isolated at the condition that
$\tau_L\lsim t_L$, {\it i.e.}, if $(q-1)d/2<2$ or equivalently $d<d_{\rm seg}$,
where $d_{\rm seg}=4/(q-1)$ is the {\it critical segregation dimension} 
announced in Eq.~(\ref{resultdseg}). Hence for $d>d_{\rm seg}$ diffusion 
prevents segregation, but for $d<d_{\rm seg}$ the reactions occur so fast that 
they segregate the system into single-species domains.

Segregation decelerates the decay of the total density. To determine precisely
how that happens, we observe that at any time $t$ the total density $\rho(t)$ 
is equal to the limit density $\rho^*A^*_{L(t)}$ in a volume of corresponding 
size, $L(t)^d$, where $L(t)\sim (t/D)^{1/2}$. Combining this with 
Eq.~(\ref{A0L}) we find that asymptotically for $t\to\infty$
$\rho(t)\sim t^{-(q-1)d/4}$, valid when segregation occurs under mean-field 
conditions. Hence in this situation we obtain the decay exponent 
$\alpha(q,d)=(q-1)d/4$ as given in Eq.~(\ref{eqnalphaqnoninteger}).

As we will see in subsection \ref{subRG}, the mean-field approximation, with or
without segregation, is valid only for $d>d_c$ (and marginally valid in 
$d=d_c$), where $d_c=2$ is the $q$-independent upper critical dimension. It 
follows that in the mean-field part of the $(q,d)$ plane the only physical 
systems ({\it i.e.,} with integer $q$ and $d$) showing segregation are the 
two-species ($q=2$) annihilation processes in $d=2$ and $d=3$ dimensions. 
We cannot exclude, however, that noninteger values of $q$ might someday be
given a physical interpretation as well.

It so happens that the $q=2$ system satisfies a microscopic conservation law: 
The difference $N_A-N_B$ between the particle numbers of the two species $A$ 
and $B$ is strictly conserved in each reaction. This law may therefore be 
viewed as being at the origin of the segregation effect. In any case, for 
$q\geq 3$ and $d>2$ we conclude that the homogeneous mean-field results of 
subsection {\it \ref{subsecmeanfield}} apply.

It is interesting to remark that the derivation of our expression 
(\ref{resultdseg}) may also be based on Ref.~\cite{benavraham86}. Here, the 
authors set $\rho_i(t)=\rho(t)+c_i(t)\sqrt{\rho^*}L^{-d/2}$, {\it i.e.,} the
time-dependent generalization of our Eq.~(\ref{rhoiL0}), and by expanding the 
master equation they show that the mean-square fluctuations 
$c(t)^2 \equiv \langle(c_i(t)-c_j(t))^2\rangle$ obey an ordinary differential 
equation in time [their Eq.~(29)]. The solution of that equation, which was 
also examined very recently in Ref.~\cite{bennaim03}, shows that for $1<q<3$ 
the typical fluctuation $c(t)\sqrt{\rho^*}L^{-d/2}$ becomes of the order of the
density $\rho(t)$ on the same time scale $\tau_L$ that appeared above in our 
study of the nonlinear solution; whence we obtain the same result for 
$d_{\rm seg}$.

\subsection{Renormalisation group arguments}
\label{subRG}

In dimension $d\leq d_c=2$ the considerations of the two preceding subsections 
{\it \ref{subsecmeanfield}} and {\it \ref{subsecGinzburg}} are no longer valid.
Fluctuations in these lower dimensions, where the reactions become effectively
diffusion-limited, need to be taken into account beyond the Gaussian 
approximation. A systematic treatment of fluctuations in reaction-diffusion
problems starts from the associated classical master equation, utilises a
standard representation in terms of bosonic creation and annihilation
operators, and therefrom builds a field theory action by means of coherent
state path integrals, see, {\it e.g.}, 
Refs.~\cite{doi76,grassberger80,peliti86,lee94}. This action is then further
analysed using dynamic renormalisation group (RG) methods, and the ensuing 
scaling laws can be extracted from the Callan-Symanzik flow equation. 

For the $q$-species MAM the ensemble average at time $t$ of any
observable $F$ can be expressed in terms of functional integrals over $2 q$
fields $\{\phi_i\}_{i=1}^q$ and $\{\hat{\phi}_i\}_{i=1}^q$ as 
\begin{equation} 
\langle F \rangle_t = \frac{\int \prod_{i=1}^q \rD \hat{\phi}_i\rD \phi_i\,
\bar{F}(\{\hat{\phi}_i\},\{\phi_i\})\, \ee^{-S[\{\hat{\phi}_i,\phi_i\}]}}
{\int \prod_{i=1}^q \rD \hat{\phi}_i\rD \phi_i\,
\ee^{-S[\{\hat{\phi_i},\phi_i\}]}} \ ,
\label{obs}
\end{equation} 
where the functional $\bar{F}(\{\hat{\phi}_i\},\{\phi_i\})$ is uniquely 
determined by the observable $F$ ({\it e.g.}, the total particle density is 
given by $\sum_{i=1}^q\hat{\phi}_i\phi_i$).
The statistical weight of any configuration is given by the action
\begin{eqnarray}
{\cal S}[\{\hat{\phi_i},\phi_i\}]&=&\sum_{i=1}^q\int \dd^d r \Biggl\{ 
\int_0^\infty \dd t \biggl[ \hat{\phi}_i (\p_t-D\nabla^2)\phi_i - \lambda
\sum_{j=1 \atop j\neq i}^q (1 - \hat{\phi}_i \hat{\phi}_j)\phi_i\phi_j \biggr] 
\nonumber\\ 
&&\qquad\qquad- \rho_i(t=0) \hat{\phi}_i(t=0) \Biggr\} \, .
\label{Sphi} 
\end{eqnarray}
The bilinear part of this action defines $q$ diffusion propagators, whereas the
reactions introduce $q(q-1)/2$ two- and four-point annihilation vertices, as 
depicted in Fig.~\ref{graphs} (a). Here, the bare coupling $\lambda$ represents
the continuum version of the uniform annihilation rate of the discrete model.
Straightforward power counting indicates that fluctuations become relevant in 
$d<2$ dimensions, for any value of $q$. Hence we identify $d_c=2$ as the upper 
critical dimension in the RG sense. Thus we need to distinguish three cases: 

\begin{figure}
\begin{center}\includegraphics[scale=0.27]{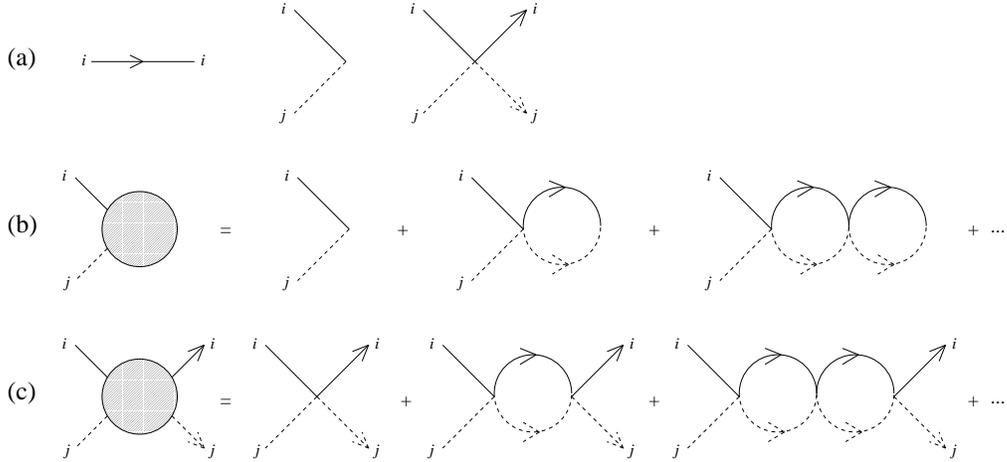}\end{center}
\caption{\label{graphs} (a) The propagators and vertices of the bosonic field 
theory. \\ (b,c) Renormalisation of the two- and four-point vertices,
respectively.}
\end{figure} 

(i) For dimensions $d>2$ the fluctuation contributions are convergent in the 
infrared, and standard perturbation theory is applicable. Therefore the 
mean-field equations (\ref{rateequations}), augmented with diffusive terms, 
should yield the correct asymptotic behaviour, whether or not species 
segregation appears. For values $q\geq 3$, the critical segregation dimension 
is $d_{\rm seg}\leq 2$, so segregation will not occur. Thus the long-time decay
of the particle densities will follow the homogeneous mean-field results 
described in subsection~{\it \ref{subsecmeanfield}}. Specifically, for equal 
initial densities $\rho_i(0) = \rho(0)/q$ the value of the decay exponent is 
universally $\alpha = 1$, independent of $q$ and $d$. For the two-species model
($q=2$), subsection~{\it \ref{subsecGinzburg}} predicts that segregation 
happens in dimensions $d<4$. As explained in more detail in 
subsection~{\it \ref{subseceffects}} below, this is in fact associated with the
local conservation of the particle density difference $\rho_1-\rho_2$, which is
a very special feature of the two-species annihilation model. For equal 
particle densities this leads to the slower decay $\rho(t) \sim t^{-d/4}$ 
\cite{ovchinnikov78,toussaint83,lee95}.

(ii) At the critical dimension $d_c=2$ (and below), the full renormalisation
group machinery is required to further analyse the action (\ref{Sphi}). As
illustrated in Fig.~\ref{graphs}(b) and (c), the Feynman diagrams associated 
with the perturbative renormalisation of the annihilation rate assume identical
structures for all $q$. Moreover, the resulting RG flow behaves as for the 
single-species annihilation process $A+A\to \varnothing$ (see 
Refs.~\cite{peliti86,lee95} for the two-species case). In two dimensions, the 
associated running renormalised coupling tends to zero logarithmically for 
large length- or timescales. Consequently, only the tree-level Feynman diagrams
involving the initial state enter the computation of the average particle 
densities, {\it cf.} subsection~{\it \ref{subseceffects}}. We may then 
immediately conclude that for $q\geq 3$, the mean-field results of 
subsection~{\it \ref{subsecmeanfield}} acquire straightforward logarithmic
corrections, independent of the value of $q$: If the density of the most
abundant particle species is nondegenerate ($n_p=1$), it will approach a 
constant at long times. The approach to this asymptotic value, as well as the
vanishing of all other species, is characterised by a {\it stretched} 
exponential $\ln [\rho_i(t)-\rho_i(\infty)]\sim -t/\ln t$. In the degenerate
case ($n_p>1$), which includes equal initial densities $\rho_i(0)=\rho(0)/q$ as
a special situation ($n_p=q$), one obtains $\rho(t) \sim t^{-1} \ln t$. Only 
the two-species MAM with equal initial densities has distinct features, with 
$\rho(t) \sim t^{-1/2}$, owing to the local conservation law and the ensuing
particle segregation (see subsection~{\it \ref{subseceffects}}). In 
Sec.~{\it \ref{secMC}}, we shall support the above predictions by means of 
Monte Carlo simulations in two and three dimensions.

(iii) In dimensions $d<2$, as explained above, the `bulk' renormalisation for 
the reaction rate proceeds precisely as for the single-species annihilation 
model, see Fig.~\ref{graphs}(b) and (c). However, this does {\it not} imply 
that the ensuing asymptotic density decay power law holds as well. For now, in 
order to compute observables such as the averages $\langle\rho_i(t)\rangle$, 
one must sum all Feynman diagrams that contain contributions from relevant 
operators on the initial time sheet ($t=0$). For the $q$-species MAM, this 
turns out to be a formidable task. Before we discuss this issue further, let us
recall the situation in the two-species annihilation model.

\subsection{Effects of the local conservation law for $q=2$}
\label{subseceffects}

The renormalisation of the annihilation rate $\lambda$ that captures the
emergence of particle anticorrelations and depletion zones remains present
(absent) in dimensions $d\leq d_c=2$ ($d>2$) and unchanged even if there exists
an additional microscopic conservation law. Yet physical effects associated 
with the latter, such as particle segregation, may crucially affect the 
system's leading asymptotic behaviour. Such phenomena, though encoded in the 
initial state, {\it viz.} the $t=0$ time sheet in the action (\ref{Sphi}), are 
in general more difficult to analyse in field theory than `bulk' 
renormalisations.

We illustrate this with a brief review of the situation for $q=2$, following
Ref.~\cite{lee95}. The two-species pair annihilation process is very special
owing to a {\it local} conservation law (not valid for any other value of $q$),
namely for the particle density difference 
$\rho_1(t)-\rho_2(t) = \rho_1(0)-\rho_2(0)$ \cite{toussaint83}. As a 
consequence, the initial state at $t = 0$ propagates into the `bulk' ($t > 0$) 
all the way to $t\to\infty$, seriously complicating the analytical approach. 
To determine the asymptotic behaviour of the density Lee and Cardy \cite{lee95}
used the following procedure. After shifting the response fields according to
$\hat{\phi_i}\to \bar{\phi}_i=\hat{\phi}_i-1$, new fields 
$\psi=(\phi_1+\phi_2) /\sqrt{2}$ and $\xi=(\phi_1-\phi_2)/\sqrt{2}$ are
introduced, along with analogous transformations for the response fields 
$\bar{\psi}$ and $\bar{\xi}$. To account for the effect of the initial state, 
specifically for equal initial densities, an effective action is built that
contains all possible initial terms compatible with the symmetries of the 
problem. These can generically be written as 
$\Delta^{(m,n)} \, \bar{\psi}^m(0)\bar{\xi}^n(0)$. The most relevant coupling 
responsible for the leading-order decay of the density turns out to be 
$\Delta=\Delta^{(0,2)}=-\Delta^{(2,0)}$, whose canonical scaling dimension is 
$4-d$. 

For dimensions $d>4$ therefore, the initial state plays no crucial role, 
and the asymptotic behaviour is determined by the homogeneous mean-field rate
equations. For $2<d\leq 4$, when the `bulk' parameters in the model do not 
require any renormalisation, a simple effective theory can be constructed. 
Aside from the new term $\Delta \, \bar{\xi}^2(0)$, it resembles the mean-field
rate equations, supplemented with diffusion terms. Since it is a locally 
conserved mode, the density difference $\xi(t)$ propagates diffusively, 
describing the growth of segregated regions in time. This already implies 
$\rho_i(t) \sim t^{-d/4}$ as $t\to\infty$ for equal initial densities. The 
amplitude of this power law is determined by summing all tree-level Feynman 
diagrams that involve the initial parameters $\rho(0)$ and $\Delta$, with the 
ultimate result \cite{oerding96,lee97}
\begin{equation}
\rho_{1,2}(t) \sim \sqrt{\frac{\rho(0)}{\pi}}\, \frac{1}{(8\pi\, D\, t)^{d/4}}
\quad \text{as} \ t\to\infty \, , \quad 2<d\leq 4 \, .
\label{solution} 
\end{equation}

In lower dimensions $d\leq 2$, however, one needs to apply the full machinery
of the renormalisation group in order to calculate the $\langle \psi \rangle$
by means of a perturbative expansion with respect to the parameter $\eps=2-d$. 
Precisely as in the single-species pair annihilation process \cite{lee94}, the
renormalised dimensionless counterpart of the reaction rate $\lambda$ flows
towards a stable fixed point of order $\eps$. Yet the calculation of 
observables in the framework of this $\eps$ expansion still necessitates 
nonperturbative summations over {\it all} orders of the initial parameters 
$\rho(0)$ and $\Delta$ \cite{lee95}. Whereas this is a straightforward task for
the single-species reaction, this goal has to date not been achieved beyond the
leading order in $\eps$ even for the two-species pair annihilation process. For
$\eps=0$, where the renormalised running coupling associated with the 
annihilation rate vanishes logarithmically, this suffices to establish the 
validity of the result (\ref{solution}) even at the upper critical dimension 
$d_c=2$. For $d<2$, though, the field-theoretic treatment in principle leaves 
three possibilities: Higher-order terms may have no crucial effect at all, 
maintaining the validity of Eq.~(\ref{solution}); they may alter the amplitude 
of the asymptotic power law, but leave the exponent $\alpha(2,d)=d/4$ intact; 
or they might even lead to a different asympotic algebraic behaviour. However, 
this last scenario is not supported by simulations in $d=1$ 
\cite{toussaint83,kang84} and exact results on a specific variant of the 
two-species annihilation model \cite{bramson91}. 

The arguments of Ref.~\cite{lee95} summarised here show that in order to 
extract the value of $\alpha(q)$ in $d=1$ within a bosonic field theory, one 
must perform a sum over all diagrams that contain relevant initial state 
operators. Even for the two-species annihilation model this summation appears 
very difficult, if not impossible. We hence expect the analysis to be even more
cumbersome for the $q$-species process. Yet the possibility of species 
segregation must be encoded precisely in these contributions from the initial
state.

To explain the $q$-dependence of the asymptotic regime, one then needs a
field theory representation where the special topology of the one-dimensional 
case becomes manifest. An example is provided by the theory of quantum spin
chains. In Appendix B, we discuss how the annihilation model can be mapped onto
an anisotropic antiferromagnetic Heisenberg spin chain. The appropriate 
continuum limit for the low-lying excitation yields generalised nonlinear sigma
models with topological Chern-Simons terms. For the $q$-species MAM, the
different topological sectors for each $q$ will become mixed through the 
annihilation processes. We conjecture that the presence of these topological
terms, effective only near one dimension, must be responsible for the 
$q$-dependence of the density decay exponent in the asymptotic regime. Yet the 
ensuing formalism is not readily amenable to a direct calculation of 
$\alpha(q,1)$ either. In view of these difficulties, we shall adopt a very 
different approach to the one-dimensional $q$-MAM in Sec.~{\it \ref{secSMAM}}. 
In higher dimensions $d\geq 2$ the topological terms do not play any 
significant role, which supports that the RG analysis of section \ref{subRG} 
predicts the correct MAM asymptotic behaviour.

\subsection{Role of special initial conditions}
\label{subsecspecialincond}

Yet before proceeding, we note that special initial conditions can alter the
generic asymptotic behaviour of the $q$-MAM. For instance, if the particles 
live on a one-dimensional lattice, their respective rank can be labeled 
starting from a given origin. Since particles only vanish in pairs, the parity 
of each particle rank is a locally conserved quantity. This immediately implies
that if one starts with a correlated configuration such as
\begin{equation*}
\ldots A_{q-1}A_qA_1A_2\ldots A_{q-1}A_qA_1A_2\ldots,
\end{equation*} 
the parity of $q$ will dictate the long-time decay of the density. On the one 
hand, if $q$ is odd, two particles of the same species always have an even 
number of particles between each other. Hence these particles can annihilate by
pairs and two particles of the same species can become nearest-neighbours after
some annihilation reactions have occurred. Therefore the subtle initial 
correlations will be wiped out eventually and segregation becomes possible, 
leading to an asymptotic density decay as in the $q$-MAM (albeit following a
potentially large crossover period). On the other hand, if $q$ is even, the 
particles lying between two identical particles cannot entirely annihilate each
other. In that situation, particles of the same species never meet on adjacent 
lattice sites, species segregation cannot occur, and the density decays as in 
the single-species pair annihilation reaction $A+A\to \varnothing$, {\it i.e.},
with $\alpha=1/2$.

\section{Simplified one-dimensional model}
\label{secSMAM}

\subsection{Simplified Mutual Annihilation Model (SMAM)}

In order to address the one-dimensional case, we now introduce a simplified 
version of the MAM, to be referred to as SMAM, which retains only the bare 
essentials of the original model. We begin by noting that the particle 
configuration may at any time be decomposed into a sequence of domains, each 
comprising only a {\it single} particle species, such that adjacent domains 
contain different species. Owing to the diffusive nature of the process the 
typical domain size increases as $\ell(t)\simeq (Dt)^{1/2}$, as we shall 
confirm numerically in Sec.~\ref{secMC}. Combining this with the expectation 
that $\rho(t)\sim (Dt)^{-\alpha}$, we may express the average particle number 
$\nav(t)=\ell(t)\rho(t)$ in a domain as
\begin{equation}
\nav(t)\,\sim\,(Dt)^{\frac{1}{2}-\alpha} \, .
\label{navrho}
\end{equation}
Clearly, unbounded domain growth, hence species segregation, can only occur if 
$\alpha<1/2$. The emergence of segregation thus necessarily implies a 
{\it slower} density decay than predicted for the homogeneous system 
($\alpha=1/2$) in one dimension.

Inspired by Eq.~(\ref{drhodt1}) we now introduce the hypothesis that 
fluctuations in the effective annihilation rate $\kappa(t)$, whether in the 
course of time or between different reaction zones, are not crucial and may be 
ignored because in one dimension the segregation effect will dominate at least
the long-time dynamics. This leads us directly to the SMAM picture: {\it The 
particle content of each domain, owing to the annihilation reactions taking 
place at both of its ends, decreases at the constant rate} $2\kappa(t)$. This 
picture is then completed by the rule for what happens when a domain 
disappears, {\it i.e.}, gets emptied of all its particles: Then, either the 
left and right neighbouring domains have the same species of particles and fuse
into a single new domain; or the neighbouring domains have different particle 
species and a new reaction zone appears between them. These two possibilities 
correspond to annihilation and coalescence of domain walls, respectively. We
already note that it is precisely via these two different scenarios, whose
respective probabilities depend on the overall number of species, that the
decay exponent $\alpha(q)$ eventually becomes a function of $q$.

The following algorithm makes this explicit. We consider a one-dimensional
lattice (`chain') whose sites, labeled by indices $i,j,\ldots$, represent the
{\it domains} of the original MAM. Below we will indifferently refer to them as
`sites' or as `domains'. Their initial number will be denoted by $N_0$. 
To start, we randomly select $N_0$ {\it number variables} 
$n_{10},n_{20},\ldots,n_{N_00}$, drawn independently from some probability 
distribution on the positive integers; these variables represent the initial 
numbers of particles in the domains. We then assign initially to each domain a 
particle type (`colour') in a random fashion, except for the restriction that
neighbouring domains are not allowed the same colour. Yet note that the
subsequent algorithm does not require that we assign these colours explicitly.
The time evolution subsequent to this random initial condition proceeds via 
{\it deterministic} iterations. The $(k+1)$th iteration changes the total 
number of sites from $N_k$ to $N_{k+1}$ and converts the set of values 
$\{n_{ik}\}_{_{i=1}}^{^{N_k}}$ to $\{n_{j,k+1}\}_{_{j=1}}^{^{N_{k+1}}}$. The 
following update consists of four steps:
\begin{enumerate}
\item The values of all $n_{ik}$ are reduced by $1$.
\item All sites $i$ which after step (i) have $n_{ik}=0$ are eliminated from 
      the lattice. The remaining sites are reconnected in such a way that their
      ordering along the chain remains unchanged.
\item Sites which as a result of step (ii) have become nearest-neighbour sites 
      and which have the same colour, are fused into a single site of that 
      colour; the number variables of the fusing sites are added together. 
\item We write $N_{k+1}$ for the remaining number of sites, and relabel these 
      sites by an index $j$ that runs from $1$ to $N_{k+1}$ along the chain.
\end{enumerate}
We began our investigation by actually running this algorithm on the computer,
and found these simulations to converge rapidly and yield accurate exponent 
values. These advantages were however superseded by our discovery that the 
above algorithm is actually amenable to an asymptotically exact solution.

In order to keep track of the colours of the sites, in our computer
implementations of this algorithm we also used a set of {\it colour variables}
$\{\sigma_{ik}\}_{i=1}^{N_k}$, where $\sigma_{ik}\in\{1,2,\ldots,q\}$ 
represents the colour of the $i$th site after the $k$th iteration. In our 
discussion below of the analytic solution of the algorithm our dealing with 
these variables will be implicit. The sole quantity we need to know is the 
probability, to be denoted $\eta_r$, that two domains separated by $r$ other 
domains have the same colour, {\it i.e.,} that $\sigma_{ik}=\sigma_{i+r+1,k}$. 
Obviously, $\eta_0=0$ and
$\eta_1=1/(q-1)$. Since the two variables $\sigma_{ik}$
and $\sigma_{i+r+1,k}$ can be equal only if $\sigma_{ik}\neq\sigma_{i+r,k}$, 
and in that case in fact coincide with probability $1/(q-1)$, it follows that 
$\eta_r$ obeys the recursion relation $\eta_r=(1-\eta_{r-1})/(q-1)$, whence 
\begin{equation}
\eta_r=[1-(1-q)^{-r}]/q \, , \quad r=0,1,2,\ldots \, .
\end{equation} 
Thus the domain colours exhibit an oscillatory short-ranged correlation. It is
easy to see that this correlation is in fact invariant under an iteration step.

After the $k$th iteration let $\bar{n}_k=N_k^{-1}\sum_{i=1}^{N_k}n_{ik}$ denote
the average number of particles per domain. Let furthermore $\rho_k$ denote the
particle density and $t_k$ the physical time after $k$ iterations. In terms of 
the data produced by the algorithm, these two quantities are given by
\begin{equation}
\frac{\rho_k}{\rho_0} = \frac{N_k\,\bar{n}_k}{N_0\,\bar{n}_0} \ , \qquad
\frac{t_k}{t_0} = \frac{N_0^2}{N_k^2} \ .
\label{densitytime}
\end{equation}
Here, the second relation holds because the typical domain length $\ell(t)$ is 
by hypothesis proportional both to the inverse total number of domains, and the
square root of time. Our purpose will be to determine $\rho_k$ and $t_k$ as 
functions of the iteration index $k$, and subsequently by elimination 
$\rho(t)$.

\subsection{SMAM evolution equation}

The key feature of the SMAM responsible for its analytic solvability is that at
every iteration $k$ the number variables $n_{ik}$, for $i=1,2,\ldots,N_k$, are 
uncorrelated. This is because they descend from disjoint sets of `ancestor' 
variables. As a result this model obeys an exact closed set of equations, which
we shall now derive and analyse. In this subsection we will suppress the 
iteration index $k$ and will denote, preceding the $k$th iteration, the number 
of domains containing precisely $n$ particles by $M_n$, the total number of 
domains by $N=\sum_{n=1}^\infty M_n$, and the fraction of domains that contain 
precisely $n$ particles by $f_n=M_n/N$. For the corresponding quantities after
the $k$th iteration we shall use primed symbols, {\it i.e.,} $M_n'$, $N'$, and 
$f'_n=M'_n/N'$. By means of appropriate combinatorics one may express $f'_n$ in
terms of the set $\{f_m\}_{m=1}^\infty$ before the iteration. We now proceed to
derive this relationship by investigating the effect of the successive steps 
(i) through (iv) of a single iteration.

In step (i) there are $M_1=Nf_1$ domains that become empty. Let
$\widetilde{M}_n$ be an auxiliary variable indicating the number of domains 
with $n$ particles (prior to any domain fusion processes), and let 
$\widetilde{N}=\sum_{n=1}^\infty\widetilde{M}_n$ denote the total number of
domains left after step (ii) of the iteration. Then
\begin{eqnarray}
\widetilde{M}_n=M_{n+1}=Nf_{n+1} \, , \quad n=1,2,3,\ldots \, , \nonumber\\
\widetilde{N}=N-N_1=N(1-f_1) \, .
\label{defNtilde}
\end{eqnarray}
Next we are interested in the domains that disappear or become created in step 
(iii). Let $\Delta N_{\rm fusion}$ denote the net change in the total number of
domains due to this process. This number is easily determined as follows: Since
the $Nf_1$ eliminated domains were randomly placed in the lattice, elementary 
statistics shows that they formed $N_{\rm s}=Nf_1(1-f_1)$ sequences of adjacent
domains. Their elimination therefore led to $N_{\rm s}$ new nearest-neighbour 
pairs. The probability $p_r$ for an eliminated sequence to contain exactly $r$ 
domains is $p_r=(1-f_1)f_1^{r-1}$, where $r=1,2,\ldots$. It follows that two 
newly created nearest-neighbour domains / sites will have equal colours with 
probability
\begin{equation}
p=\sum_{r=1}^\infty p_r\eta_r=\frac{1}{q-1+f_1} \ .
\label{eqnp}
\end{equation}
We remark in passing that whereas in our earlier paper \cite{deloubriere02} the
possibility of $r$ domains disappearing simultaneously was correctly taken into
account, the exact expression (\ref{eqnp}) for $p$ was 
approximated by its low-density, 
long-time limiting value $p=1/(q-1)$. 
The exact expression (\ref{eqnp}) used here
actually leads to simpler mathematics, and of course does not change 
the leading power laws; its final consequence is a different amplitude in the 
subleading term of the end result, Eq.~(\ref{rhoasptt}). 

Since the total number of new nearest-neighbour pairs is $N_{\rm s}$, the 
change $\Delta N_{\rm fusion}$ of the total number of domains due to fusion in 
step (iii) is 
\begin{equation}
\Delta N_{\rm fusion}=-N_{\rm s}p=-N\,\frac{f_1(1-f_1)}{q-1+f_1} \ .
\label{eqnDeltaNfusion}
\end{equation}
Eqs.~(\ref{defNtilde}) and (\ref{eqnDeltaNfusion}) allow us to directly relate 
$N'$ to $N$ according to
\begin{equation}
N'=\widetilde{N}+\Delta N_{\rm fusion}=N\,\frac{(q-1)(1-f_1)}{q-1+f_1} \ .
\label{fullNprimeN}
\end{equation}
However, we have to study step (iii) in greater detail. For we need to know the
numbers $\Delta\widetilde{M}_{n,\rm loss}$ and 
$\Delta\widetilde{M}_{n,\rm gain}$ of domains of size $n$ that are destroyed 
and created, respectively, during that step. Consider, after step (ii), a 
specific domain (with arbitrary particle number $n$). This very domain will 
disappear owing to fusion if its left {\it or} its right neighbour are of the 
same colour. The probability $p_{\rm c}$ that one given neighbour is of the 
same colour is $p_{\rm c}=f_1/(q-1+f_1)$, namely the product of the probability
$f_1$ of its originally being adjacent to a group of domains of size $1$, with
the probability $1/(q-1+f_1)$ of this group being adjacent on its other end to 
a domain of the same colour as the specific one we are considering. Since it 
has two neighbours, the given domain under consideration will disappear with 
probability $2p_{\rm c}-p_{\rm c}^2$. Since there are altogether 
$\widetilde{M}_n=Nf_{n+1}$ domains of size $n$, we find
\begin{eqnarray}
\Delta\widetilde{M}_{n,\rm loss}&=&-\widetilde{M}_n(2p_{\rm c}-p_{\rm c}^2)
\nonumber\\
&=& -Nf_{n+1} \left[1-\left(\frac{q-1}{q-1+f_1}\right)^2\right] , \quad
n=1,2,\ldots \, .
\label{eqnDeltaNnloss}
\end{eqnarray}
Summing this result over $n=1,2,\ldots$ yields the total number 
$\Delta\widetilde{N}_{\rm loss}$ of domains that disappear in step (iii),
\begin{equation}
\Delta\widetilde{N}_{\rm loss}=-N(1-f_1)\left[1-\left(\frac{q-1}{q-1+f_1}
\right)^2\right] .
\label{eqnDeltaNloss}
\end{equation}

Next we turn to the number $\Delta\widetilde{M}_{n,\rm gain}$ of $n$-particle 
domains created in step (iii). Since such a creation may result from the fusion
of $s$ domains, we need to determine the corresponding number for each value of
$s$, and subsequently sum over all $s=1,2,\ldots.$ After step (ii) there are 
$\widetilde{N}=N(1-f_1)$ domains. We may therefore write
\begin{equation}
\Delta\widetilde{M}_{n,\rm gain}=N(1-f_1)\sum_{s=2}^\infty x_s R_{sn} \, , 
\quad n=1,2,\ldots \, .
\label{eqnDeltaNngain}
\end{equation}
Here $x_s$ represents the probability that a specific domain undergoes a fusion
with the $s-1$ domains that follow along the chain, and $R_{sn}$ is the 
independent probability that this fusion produces a domain of size $n$. We now 
determine $x_s$ and $R_{sn}$. In order that exactly $s$ domains fuse, they must
subsequent to step (ii) form a sequence of adjacent domains of the same colour,
bordered at both ends by domains of a different colour. The probability for 
this arrangement is 
\begin{equation}
x_s=(1-p_{\rm c})^2p_{\rm c}^{s-1} \, .
\label{eqnxs}
\end{equation}
The domain fractions after step (ii) are 
$\widetilde{f}_n=\widetilde{M}_n/\widetilde{N}=f_{n+1}/(1-f_1)$, with
$n=1,2,\ldots$. The probability that the $s$ fusing domains have sizes
$m_1,m_2,\ldots,m_s$ is equal to the product 
$\widetilde{f}_{m_1}\widetilde{f}_{m_2}\ldots\widetilde{f}_{m_s}$, and the 
probability that this fusion produces a domain of size $n$ is in turn equal to 
this product summed on all sizes of the fusing domains, subject to the
constraint $m_1+m_2+\ldots+m_s=n$. Hence we have
\begin{equation}
R_{sn}=(1-f_1)^{-s}\sum_{m_1=1}^\infty\ldots\sum_{m_s=1}^\infty
f_{m_1+1}\ldots f_{m_s+1}\,\delta_{n,m_1+\ldots+m_s} \ ,
\label{eqnysn}
\end{equation}
Using the preceding expressions in Eq.~(\ref{eqnDeltaNngain}) and
summing over $n=1,2,\ldots$ yields
\begin{equation}
\Delta\widetilde{N}_{\rm gain}=N\,\frac{f_1(1-f_1)(q-1)}{(q-1+f_1)^2} \ .
\label{eqnDeltaNgain}
\end{equation}
From Eqs.~(\ref{eqnDeltaNfusion}), (\ref{eqnDeltaNloss}), 
and (\ref{eqnDeltaNgain}) one may check that $\Delta N_{\rm fusion}=
\Delta\widetilde{N}_{\rm gain}+\Delta\widetilde{N}_{\rm loss}$, as must of
course be the case.

We now consider the recursion
\begin{equation}
M'_n=\widetilde{M}_n+\Delta\widetilde{M}_{n,\rm loss}+
\Delta\widetilde{M}_{n,\rm gain} \, , \quad n=1,2,\ldots \, ,
\end{equation} 
and divide both sides of this equation by $N'$ so that the right-hand side 
becomes equal to the fraction $f'_n$. With the aid of several of the preceding 
equations the right-hand side may then be expressed explicitly in terms of the 
original fractions $\{f_m\}_{m=1}^\infty$. One thereby finds the SMAM 
{\it evolution equation}
\begin{equation}
f'_n=\frac{q-1}{(1-f_1)(q-1+f_1)}\,{\cal R}_n \, ,
\label{fullrecursion}
\end{equation}
valid for $n=1,2,\ldots$, with
\begin{equation}
{\cal R}_n=\sum_{s=1}^\infty g_1^{s-1} \sum_{m_1=1}^\infty\ldots
\sum_{m_s=1}^\infty f_{m_1+1}\ldots f_{m_s+1}\,\delta_{n,m_1+\ldots+m_s} \ ,
\label{defRn}
\end{equation}
and where we have introduced the abbreviation
\begin{equation}
g_1=\frac{f_1}{(1-f_1)(q-1+f_1)} \ .
\label{defg1}
\end{equation}
Eq.~(\ref{fullrecursion}), together with (\ref{defRn}) and (\ref{defg1}), 
constitutes an explicit and fully exact recursion relation for the evolution of
the domain fractions under the deterministic SMAM algorithm. One easily checks 
that it conserves the sum rule $\sum_{n=1}^\infty f_n=1$. Recall that the term 
with index $s$ in the sum in Eq.~(\ref{defRn}) represents the creation of a 
domain of $n$ particles by the simultaneous fusion of $s$ domains. The terms
with $s\geq 3$ are obviously quite model-specific and one would expect the 
essential physics to be embodied already in the lowest-order nonlinearity, 
{\it i.e.}, in the contribution with $s=2$. However, in spite of the cumbersome
appearance of the above recursion, its mathematical analysis turns out to be 
easier when all terms are retained.

\subsection{Asymptotic density decay}

We now restore the iteration indices $k$ and $k+1$ to replace the unprimed and
primed variables, respectively, of the previous subsection. In order to find a 
solution to the SMAM evolution equation (\ref{fullrecursion})-(\ref{defg1}) we 
substitute as an ansatz an exponential distribution 
$f_{nk}=\epsilon_k(1-\epsilon_k)^{n-1}$ into the right-hand side of
Eq.~(\ref{fullrecursion}). The recursion then reproduces an exponential 
distribution, $f_{n,k+1}=\epsilon_{k+1}(1-\epsilon_{k+1})^{n-1}$, with
\begin{equation}
\epsilon_{k+1}=\frac{q-1}{q-1+\epsilon_k}\,\epsilon_k \, .
\label{epsrecursion}
\end{equation}
Hence we have indeed identified a solution. For this solution we may express
Eq.~(\ref{fullNprimeN}) as
\begin{equation}
N_{k+1}=\frac{(q-1)(1-\epsilon_k)}{q-1+\epsilon_k}\,N_k \, .
\label{Nrecursion}
\end{equation}
In view of Eq.~(\ref{densitytime}) and the fact that $\nav_k=1/\epsilon_k$,
dividing Eq.~(\ref{Nrecursion}) by (\ref{epsrecursion}) gives the recursion 
relation for the particle density $\rho_k$,
\begin{equation}  
\rho_{k+1}=(1-\epsilon_k)\rho_k \, .
\label{rhorecursion}
\end{equation}
The random initial distribution that we are considering determines the initial 
condition for Eq.~(\ref{epsrecursion}), {\it viz.} $\epsilon_0=(q-1)/q$, so
$f_{n0}=(q-1)/q^n$. 

Solving Eqs.~(\ref{epsrecursion}) and (\ref{rhorecursion}) explicitly now
appears to be quite simple. From Eq.~(\ref{epsrecursion}), with the prescribed 
initial condition, we find 
\begin{equation}
\epsilon_k=\frac{q-1}{q+k} \ ,
\label{soleps}
\end{equation}
whence, after substitution of this result in Eq.~(\ref{rhorecursion}),
\begin{equation}
\frac{\rho_k}{\rho_0}=\frac{\Gamma(q)\,\Gamma(k+1)}{\Gamma(k+q)} \ .
\label{solrho}
\end{equation} 
Next we may determine $t_k$ with the aid of Eq.~(\ref{densitytime}) and the 
fact that $\nav_k=1/\epsilon_k$,
\begin{equation}
\frac{t_k}{t_0}=\left(\frac{\rho_0\,\epsilon_0}{\rho_k\,\epsilon_k}\right)^2
=\left[\frac{\Gamma(k+q+1)}{\Gamma(q+1)\,\Gamma(k+1)}\right]^2 .
\label{solt}
\end{equation}
The desired time-dependence of the particle density $\rho(t)$ is then obtained 
by elimination of the iteration index $k$ from Eqs.~(\ref{solrho}) and 
(\ref{solt}) for $\rho_k$ and $t_k$. We will be able to carry this out 
explicitly only in an asymptotic expansion for large $k$. For $k \gg q$, one
easily finds from Eqs.~(\ref{solrho}) and (\ref{solt}) that
\begin{eqnarray}
\frac{\rho_k}{\rho_0}&=&\frac{\Gamma(q)}{k^{q-1}} \left( 1-\,\frac{(q-1)q}{2k}
\,+\ldots\right) , 
\label{rhoasptk}\\
\frac{t_k}{t_0}&=&\frac{k^{2q}}{[\Gamma(q+1)]^2} \left( 1+\,\frac{q(q+1)}{k}
\,+\ldots\right).
\label{epsasptk}
\end{eqnarray}
Setting $\tau=Ct_k=[\Gamma(q+1)]^2 t_k/t_0$ and inverting the expansion 
(\ref{epsasptk}) gives
\begin{equation}
k=\tau^{1/2q}\left( 1-\frac{q+1}{2}\, \tau^{-1/2q}+\ldots \right) .
\label{kasptt}
\end{equation}
Finally, substituting Eq.~(\ref{kasptt}) in (\ref{rhoasptk}) and identifying
$\rho_k=\rho(t_k)$ yields
\begin{equation}
\rho(t)\simeq\rho_0\,\Gamma(q)\left[ (Ct)^{-(q-1)/(2q)} + \frac{q-1}{2} \,
(Ct)^{-1/2}\right] ,
\label{rhoasptt}
\end{equation}
{\it i.e.}, the leading-order term of Eq.~(\ref{rhoasptt}) describes a 
power-law density decay $\rho(t)\sim t^{-\alpha(q)}$ with the exponent
(\ref{eqnalphaq1}); since $\alpha(q)<1/2$ we conclude that segregation occurs 
for all $2\leq q<\infty$. The correction term in Eq.~(\ref{rhoasptt}) decays 
with the power $1/2$, which corresponds to the single-species decay law in the
absence of segregation. Notice that indeed the relative amplitude of this term
grows with $q$. In field theory language, it represents the effect of reaction
rate renormalisation (without segregation) due to particle depletion. We note
that for large $q$ it will become difficult to numerically distinguish this 
correction from the leading-order term. 

It is not certain that the correction terms that we have identified here for 
the SMAM have the same relevance for the original MAM as we believe the 
leading-order term does. Nevertheless, on physical grounds one would expect the
presence of a contribution $\sim t^{-1/2}$. Indeed, as we shall see in 
Sec.~\ref{secMC}, Monte Carlo simulation estimates of the MAM exponents
$\alpha(q)$ in one dimension yield systematically {\it larger} values than our 
prediction (\ref{eqnalphaq1}). We interpret this deviation as the influence of
an additional $t^{-1/2}$ term, which should also be responsible for the 
discrepancy between the present theory and earlier simulation results 
\cite{benavraham86,krapivsky00}. In fact, the detailed analysis of their Monte
Carlo data, without prior knowledge of our work, led the authors of 
Ref.~\cite{zhong03} to the conclusion that the MAM was best described by a 
superposition of the two power laws present in Eq.~(\ref{rhoasptt}).

\subsection{Reaction zone width}

Let $\ell_{\rm int}$ be the typical interparticle distance between two unequal 
particles. In one dimension this is also the width of the reaction zone, 
hypothesised to grow with time as $t^{\lambda_r(q)}$. 
Following the arguments of
Ref.~\cite{leyvraz92} as exposed in Sec.~{\it \ref{ann}}, one is directly led 
to Eq.~(\ref{expsca}), {\it viz.} 
$\lambda_r(q)=(2q-1)/(4q)$. For $q=2$ this yields 
the result $\lambda_r(2)=3/8$ derived in Ref.~\cite{leyvraz92} for the 
two-species system. For $q=\infty$ one obtains $\lambda(\infty)=1/2$, which 
means that the reaction zone width grows as fast as the typical domain size.
This is possible only if the domains are made up of at most a finite number of 
particles, in accord with the disappearance of segregation as $q\to\infty$ and 
the recovery of $\alpha(\infty)=1/2$ as for the single-species system.

\subsection{Connection to the $q$-state Potts model}

In the continuum limit colliding domain walls coalesce to a single one with
probability $(q-2)/(q-1)$ and annihilate with the complementary probability
$1/(q-1)$. This is precisely what happens to domain walls in the dynamics of 
the one-dimensional zero-temperature $q$-state Potts model. In that context, 
one may ask for the probability $P_2(t)$ that after a time $t$ a domain wall 
that was initially present has not yet disappeared or coalesced with another 
domain wall \cite{majumdar}. In the Potts model one sets 
$P_2(t)\sim t^{-\theta(q)}$, with a $q$-dependent persistence exponent 
$\theta$. Although it is known that $\theta(2)=1/2$ and $\theta(\infty)=3/2$, 
so far an explicit expression for the function $\theta(q)$ has not been found
(see Ref.~\cite{odonoghue02}). 

The same question may be asked for the SMAM. Within this simplified model it is
easy to find the answer. The calculation is most easily done in the continuum
limit. Here we state only the result, {\it viz.} $\theta(q)=1-1/q$, which is 
in fact identical to the result obtained by Majumdar and Cornell 
\cite{majumdar98} in a mean-field approximation to the $q$-state Potts model. 
We feel this agreement is not a coincidence, but due to the fact that the 
nonfluctuating rate hypothesis at the basis of the SMAM actually represents a
kind of mean-field approximation as well. We conjecture that if appropriate 
fluctuation terms were added to the SMAM evolution equation, then $\alpha(q)$ 
would remain unchanged, whereas $\theta(q)$ would acquire nontrivial 
corrections and deviate from its mean-field value.

\section{Monte Carlo simulations}
\label{secMC}

In this section, we report our Monte Carlo simulation results which aim, within
the numerical accuracy at our disposal, to support the above analytical results
concerning the exponent $\alpha(q,d)$ in dimensions $d = 1$, $2$, and $3$. We
begin with a brief description of our simulation algorithm.

First, as appropriate data structure we chose to implement an array-based 
pseudo-binary tree to minimise the time complexity of the following four 
procedures: selecting a random entry, adding new entries, removing entries, and
retrieving the number of entry matches within the database. In terms of the 
particle reactions, these processes respectively correspond to selecting a 
random particle, adding new particles at a given lattice site, removing 
particles from a given site, and retrieving the number of particles at a given 
site. This data structure proved to be efficient in single-species reactions. 
Therefore, it was a natural choice to extend its use to multiple particle 
species. This was accomplished by maintaining separate lists of the particle 
positions for each species; that is, one instance of the tree data structure 
would store all $A$ particles, one instance of the tree data structure would 
store all $B$ particles, etc. As a consequence of this data structure, dropping
site exclusion constraints ({\it i.e.}, allowing only at most one particle per 
site) requires no extra computational effort, and we also initially allow 
particles of different species to coexist on the same site. As the simulation 
process is sufficiently robust to avoid any systematic error in the form of 
reaction biases, and since by diffusion and reaction processes, the probability
of multiple occupancy for any site diminishes rapidly in time, the asymptotic
long-time power laws should not be significantly affected by the chosen initial
conditions (with the exception of very special {\it correlated} initial states
in one dimension, as will be discussed below).

Second, the subsequent simulation steps for our implementation of the MAM 
reactions in $d$ dimensions with $q$ reacting species are as follows:
\begin{enumerate}

\item	$q$ (not necessarily empty) sites are selected at random and filled 
        with a particle such that exactly one particle of each species is added
        to the lattice and only one particle is added to any of the $q$ sites.

\item	Repeat step (i) until the lattice is filled to the desired initial 
	density, usually $1$.

\item	Increment time by $1/N(t)$ where $N(t)$ is the total number of 
	particles at this step ({\it i.e.}, apply an asynchronous time update 
	\cite{privman97,hinrichsen00}).

\item	Pick a random particle, giving equal weight to all particles;
        more precisely: \\
	(a) first select the species randomly, weighted by the ratio of the 
	number of particle of the given species to the total particle number;\\
	(b) next, randomly select a particle from this species list, giving 
	equal weight to all particles of this species.

\item	Pick a site randomly from the set of the $2d$ nearest neighbours of 
	the selected particle, observing periodic boundary conditions.

\item	If the chosen neighbouring site has no particles of a species 
	different than the selected particle, the selected particle simply hops
	from its original site to the neighbouring site.

\item 	If the chosen neighbouring site contains particles of a species 
	different than the selected particle, then one of these particles is 
	randomly selected to be removed along with the original particle.

\item 	Repeat steps (iii) through (vii) until no particles are left or until 
	the overall allotted simulation time has elapsed.

\end{enumerate}

In the following, we will present our simulation results for the $q$-species
MAM with $q=2$, $3$, $4$, and $5$. The two-species pair annihilation process 
($q=2$) was investigated numerically rather intensively in one 
\cite{toussaint83,kang84}, two \cite{toussaint83,cornell92} and three
dimensions \cite{leyvraz92}. For comparison, we will also show the 
corresponding data for the single-species annihilation process 
$A + A \to \varnothing$. As argued before, this process should describe the 
$q \to \infty$ limit of the $q$-species MAM in any dimension. Recently,  
Zhong, Dawkins, and ben-Avraham have studied the one-dimensional MAM for $q=3$,
$4$, and $5$ by means of a renormalised reaction-cell method for systems that
are equivalent to lattices with up to $2^{28}$ (!) sites \cite{zhong03}.

\subsection{Segregation in the one-dimensional MAM}

We begin with the MAM in one dimension, where our previous analysis predicts
species segregation to occur for generic (random) initial placements of the
particles along the chain. The Monte Carlo simulations were performed on a 
lattice of $10^5$ sites with periodic boundary conditions, and our data sets 
were averaged over $50$ independent runs. For each value of $q$, the long-time
density decay is indeed characterised by different exponents $\alpha(q,1)$, see
Refs.~\cite{deloubriere02,zhong03}. In order to avoid any dependence on the
details of the fitting algorithm and also probe the presence of crossover 
regimes, we follow the standard procedure to measure and plot the `local' (in
time) effective exponent $\alpha(t)$ vs. Monte Carlo time $t$ as defined 
through
\begin{equation}
\alpha(t) = - \frac{\dd \ln \rho(t)}{\dd \ln t} \ ,
\end{equation}
which in the limit $t \to \infty$ should approach $\alpha(q,1)$. 

\begin{figure}
\begin{center}
\includegraphics[scale=0.46]{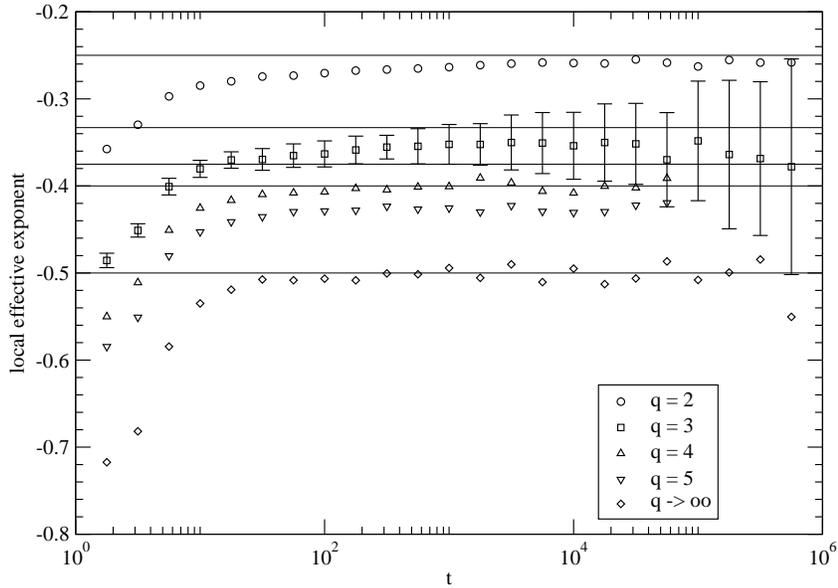}
\caption{\label{1dMAM} Local effective decay exponent $- \alpha(t)$ for the 
one-dimensional MAM with $q=2$, $3$, $4$, and $5$ (random initial conditions, 
equal initial particle numbers for each species). The data were obtained from
$10^5$ lattice sites, and are averaged over $50$ runs. For the $q=3$ data,
we also display the statistical error bars; similar accuracies apply to the 
other graphs as well. The straight lines correspond to the values 
$\alpha(q,1) = (q-1)/(2q)$ as asymptotically predicted by the SMAM. For 
comparison, results for the single-species annihilation process 
$A + A \to \varnothing$ are displayed also, corresponding to the limit 
$q \to \infty$.}
\end{center}
\end{figure}

\begin{figure}
\begin{center}
\includegraphics[scale=0.46]{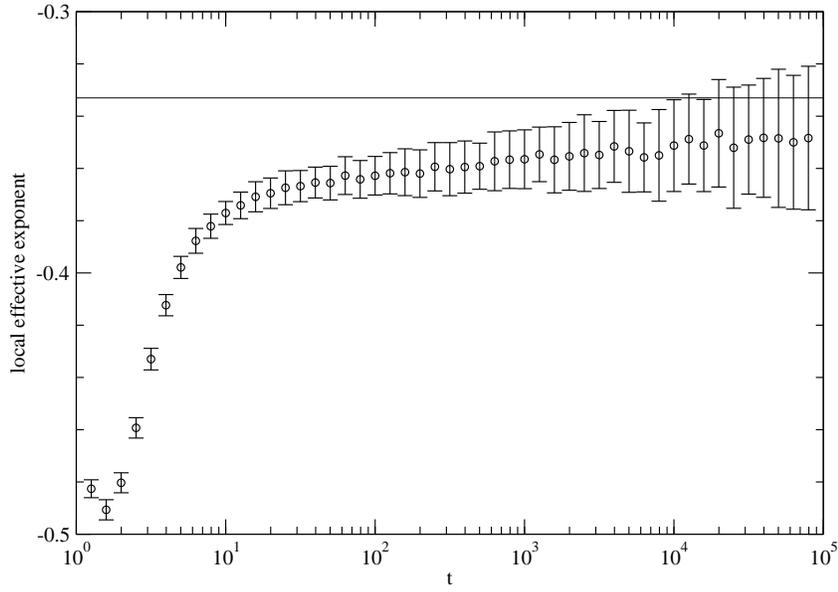}
\caption{\label{q3MAM} Local effective decay exponent $- \alpha(t)$ for the 
one-dimensional MAM with three particle species ($q=3$) with random initial 
conditions, obtained from averaging over $100$ runs on $10^6$ lattice sites.
The straight line corresponds to the expected value $\alpha(3,1) = 1/3$.}
\end{center}

\end{figure}
\begin{figure}
\begin{center}
\includegraphics[scale=0.46]{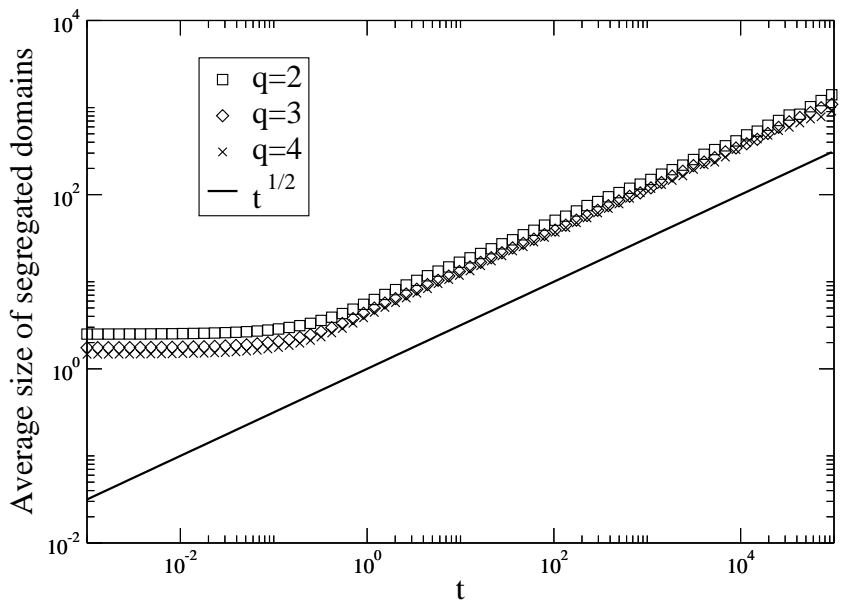}
\caption{\label{1ddomains} One-dimensional MAM for $q=2$, $3$, and $4$. The 
average size of single-species segregated regions grows as $t^{1/2}$.}
\end{center}
\end{figure}

Our data as obtained for the one-dimensional MAM with $q=2$, $3$, $4$, and $5$ 
as well as for the single-species pair annihilation (corresponding to the limit
$q \to \infty$) are depicted in Fig.~\ref{1dMAM}. The horizontal lines indicate
the SMAM predictions $\alpha(2,1) = 1/4$, $\alpha(3,1) = 1/3$, 
$\alpha(4,1) = 3/8$, $\alpha(5,1) = 2/5$, and $\alpha(\infty,1) = 1/2$ (from 
top to bottom). While the single-species data agree well with the theoretical
prediction, for finite $q$ our local effective decay exponent values in the
accessible time window are systematically lower than the asymptotic SMAM 
numbers. We interpret these data to reflect the strong next-to-leading
corrections $\sim t^{-1/2}$ to the leading density power law decay, {\it cf.} 
Eq.~(\ref{rhoasptt}). This assertion is in fact borne out by the careful data
analysis carried out by Zhong {\it et al.} for their exceedingly large systems
equivalent to lattices of up to $2^{28}$ sites. Yet even then marked 
corrections to scaling were still prominent. Notice that the simulation data 
for our system size become rather unreliable for $t > 10^5$. In 
Fig.~\ref{q3MAM} we show simulation results for the three-species MAM ($q=3$) 
obtained on a larger lattice with $10^6$ sites, averaged over $100$ runs. The 
data appear to systematically but slowly approach the expected decay exponent 
value $\alpha(3,1) = 1/3$ at long times.

In addition, we tested a crucial input for our analysis of the SMAM, namely 
that the size of the segregated domains was assumed to grow with time 
$\sim t^{1/2}$. Neglecting the effect on the asymptotic regime of the single 
particle domains, one can look at only one realisation, averaging the domain 
size over all domains in the system to obtain almost noiseless data, as shown 
in Fig.~\ref{1ddomains} for $q=2$, $3$, and $4$. The measured domain growth 
power law confirms our hypothesis and therefore supports our claim that in the 
long-time regime the SMAM and the MAM become equivalent.

We also investigated the role of special, highly correlated initial conditions
on the long-time MAM evolution. As explained in Sec.~\ref{subsecspecialincond},
the separation of distinct species in an ordered alternating alignment 
$\ldots A_{q-1}A_qA_1A_2 \ldots A_{q-1}A_qA_1A_2 \ldots$ will be preserved 
under the MAM kinetics provided $q$ is even, whence the single-species decay 
with $\alpha = 1/2$ should be recovered. For odd $q$, on the other hand, the
initial correlations disappear after some crossover period, and eventually
segregation will occur, with the density decay governed by the exponents
$\alpha(q,1)$. These predictions are confirmed by our simulation data for $q=2$
(on $65536$ sites), $q=3$ (on $98304$ sites), and $q=4$ (on $65536$ sites), 
each averaged over $20$ runs.

\begin{figure}
\begin{center}
\includegraphics[scale=0.46]{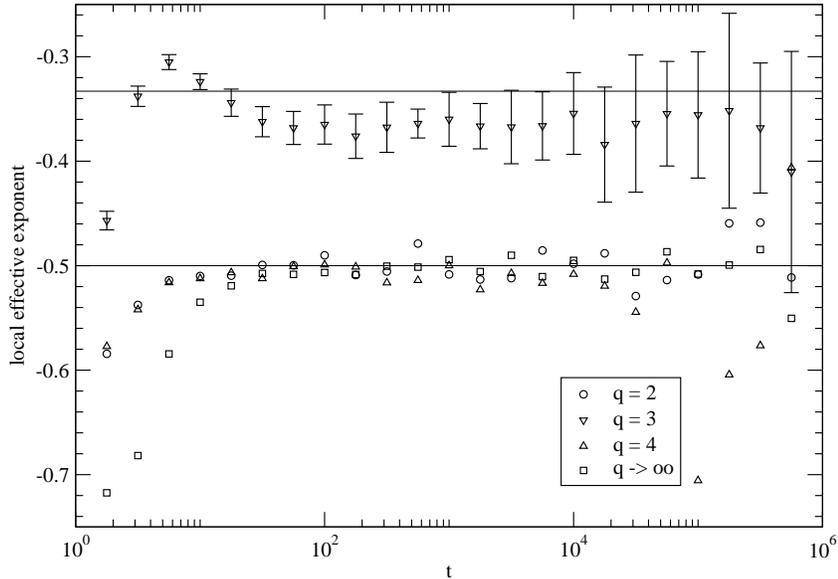}
\caption{\label{1sMAM} Local effective decay exponent $- \alpha(t)$ for the 
one-dimensional MAM with alternating initial particle placement for $q=2$ 
($\ldots ABABAB \ldots$ on $65536$ sites), $q=3$ ($\ldots ABCABC \ldots$ on 
$98304$ sites), and $q=4$ ($\ldots ABCDABCD \ldots$ on $65536$ sites), each 
averaged over $20$ runs. For the $q=3$ data, we again display the statistical 
error bars. The straight lines correspond to the theoretical expectations: 
segregation as for random initial conditions only occurs for odd $q$, while for
even $q$ such correlated initial states lead to the single-species decay 
$\sim t^{-1/2}$.}
\end{center}
\end{figure}

\subsection{Simulation results for the MAM in $d = 2$ and $d = 3$ dimensions}

\begin{figure}
\begin{center}
\includegraphics[scale=0.46]{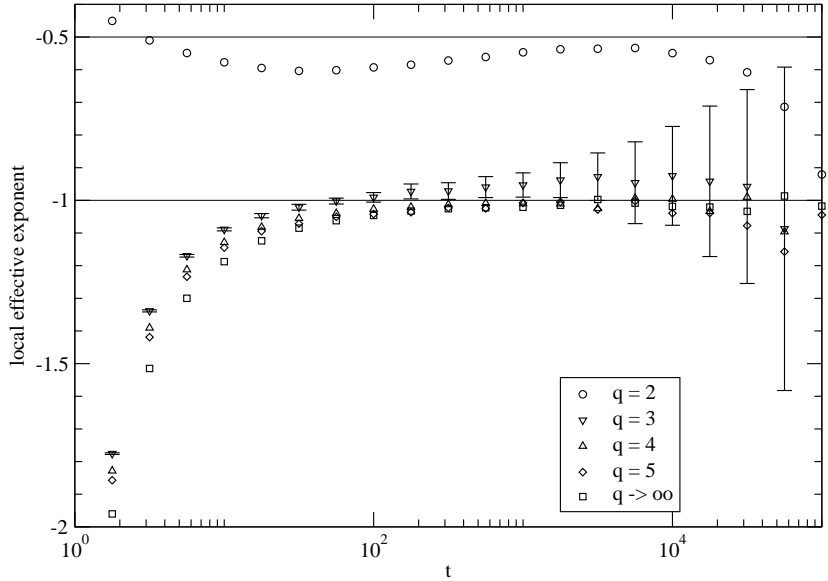}
\caption{\label{2dMAM} Local effective decay exponent for the MAM in $d=2$
dimensions for $q=2$, $3$, $4$, and $5$ (random initial conditions, equal 
initial particle numbers for each species). For $q \geq 3$, the effective
exponent $\bar{\alpha}(t)$ is obtained from $\rho(t) / \ln t$ rather than the 
density itself. The data were obtained on a $1000 \times 1000$ square lattice, 
and are averaged over $50$ runs. For the $q=3$ data, we display the statistical
error bars. The straight lines correspond to the theoretical expectations, 
$\rho(t) \sim t^{-1/2}$ for the two-species case, and 
$\rho(t) \sim t^{-1} \ln t$ for $q \geq 3$.}  
\end{center}
\end{figure}

For our two-dimensional simulations, we used a $1000 \times 1000$ sites square 
lattice with periodic boundary conditions in each direction. We have argued 
above that the single-species asymptotic decay law $\rho(t) \sim t^{-1}\ln t$ 
should apply for all $q \geq 3$ here, in contrast with the two-species process 
for which $\alpha(2,2) = 1/2$ as a consequence of the local conservation law 
for the particle density difference $\rho_1-\rho_2$. In order to test these 
assertions, we plot the local effective density decay exponent $\alpha(1)$ for
$q = 2$, whereas for $q \geq 3$ we define
\begin{equation}
\bar{\alpha}(t) = - \frac{\dd \ln \left( \rho(t) / \ln t \right)}{\dd\ln t} \ .
\end{equation}
These quantities are plotted in Fig.~\ref{2dMAM} for $q=2$, $3$, $4$, $5$ and
$\infty$ (obtained from the single-species reaction), from data averaged over 
$50$ runs. As expected $\bar{\alpha}(t) \to 1$ as $t\to\infty$, which confirms
the universal density decay for $d=2$ and hence the absence of species 
segregation, clearly distinct from the behaviour for $q=2$. We consider the
data reliable up to $t = 10^4$.

\begin{figure}
\begin{center}
\includegraphics[scale=0.46]{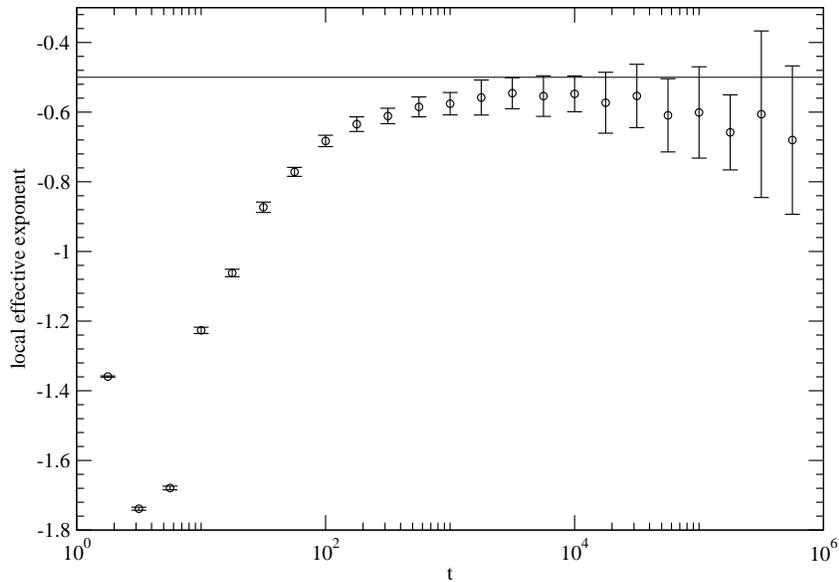}
\caption{\label{2sMAM} Local effective decay exponent $- \alpha(t)$ for the 
two-species annihilation process in $d=2$ with checkerboard initial 
state on $2048 \times 2048$ sites, averaged over $20$ runs (with statistical
error bars). Following an initial crossover period, this correlated initial 
state disappears and segregation ensues with $\rho(t) \sim t^{-1/2}$ at long 
times.}
\end{center}
\end{figure}

Again, the role of correlated initial states can be investigated. To this end,
we ran simulations for the two-dimensional two-species annihilation process 
with checkerboard initial particle placement (alternating 
$\ldots ABABAB \ldots$ rows) on fairly large square lattices with
$2048 \times 2048$ sites, averaging the data over $20$ runs. In contrast to its
one-dimensional counterpart, the stochastic reaction kinetics destroys the
initial correlations, whence after some crossover period particle segregation
emerges and consequently $\rho(t) \sim t^{-1/2}$ at long times as for random
initial configurations.

\begin{figure}
\begin{center}
\includegraphics[scale=0.46]{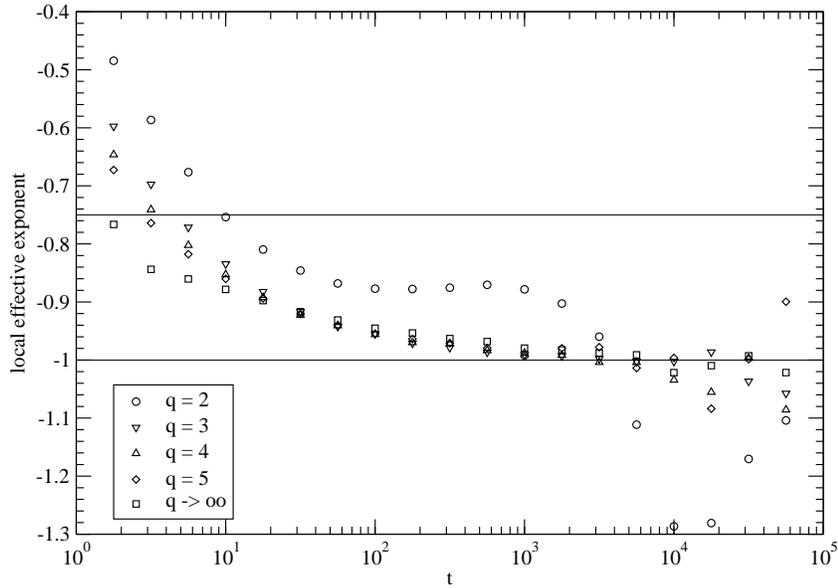}
\caption{\label{3dMAM} Local effective decay exponent $- \alpha(t)$ for the MAM
in $d=3$ dimensions for $q=2$, $3$, $4$, and $5$ (random initial conditions, 
equal initial particle numbers for each species). The data were obtained from
a $144 \times 144 \times 144$ cubic lattice, and are averaged over $50$ runs. 
The straight lines correspond to the values theoretical expectations, 
$\rho(t) \sim t^{-3/4}$ for the two-species case, and $\rho(t) \sim t^{-1}$ for
$q \geq 3$.}  
\end{center}
\end{figure}

In three dimensions, the simulations were performed for $q=2$, $3$, $4$, $5$, 
and $\infty$ on a $144 \times 144 \times 144$ cubic lattice, and the data 
averaged over $50$ runs. As anticipated above, we observe in Fig.~\ref{3dMAM} 
the MAM mean-field behaviour with $\alpha=1$ for $q \geq 3$, which demonstrates
again the absence of both reaction rate renormalisation and species segregation
in the system. In contrast, the two-species model should still display particle
segregation and follow a slower decay with $\alpha(3,2) = 3/4$. Notice that
while our data for $q = 2$ clearly show behaviour quite distinct from the cases
with $q \geq 3$, our numerical results are markedly off from the expected 
asymptotic exponent, probably again as a consequence of the competing 
$t^{-1}$ power law. Beyond $t=10^3$ the effect of the lattice periodicity sets 
in and our data are no longer useful for extracting the exponent $\alpha$.

To summarise, our simulation data confirm the universal $q$-independent 
particle density decay laws for $q \geq 3$ in dimensions $d \geq 2$. It is only
due to the presence of an extra and special conservation law that the 
two-species model behaves qualitatively differently. Correlated initial 
conditions do not appear to play a crucial role in dimensions $d \geq 2$.

\subsection{Simulation results for two- and three-lane systems}

\begin{figure}
\begin{center}
\includegraphics[scale=0.46]{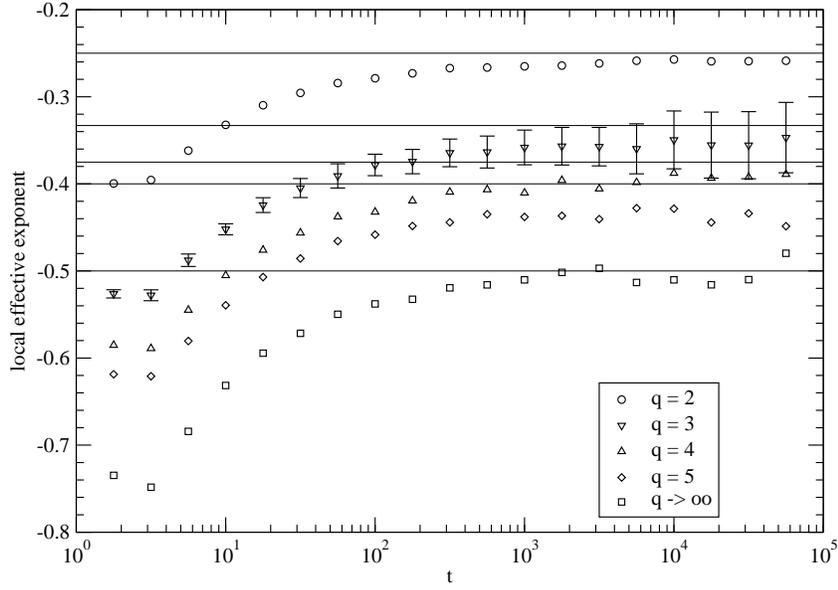}
\caption{\label{2lMAM} Local effective decay exponent $-\alpha(t)$ for the MAM 
on a two-lane system for $q=2$, $3$, $4$, and $5$ (random initial conditions, 
equal initial particle numbers for each species). The data were obtained on 
$2 \times 100000$ lattice sites, and are averaged over $50$ runs. For the $q=3$
data, we display the statistical error bars.The straight lines correspond to 
the values $\alpha(q,1) = (q-1)/(2q)$ as asymptotically predicted by the 
one-dimensional SMAM.}
\end{center}
\end{figure}

\begin{figure}
\begin{center}
\includegraphics[scale=0.46]{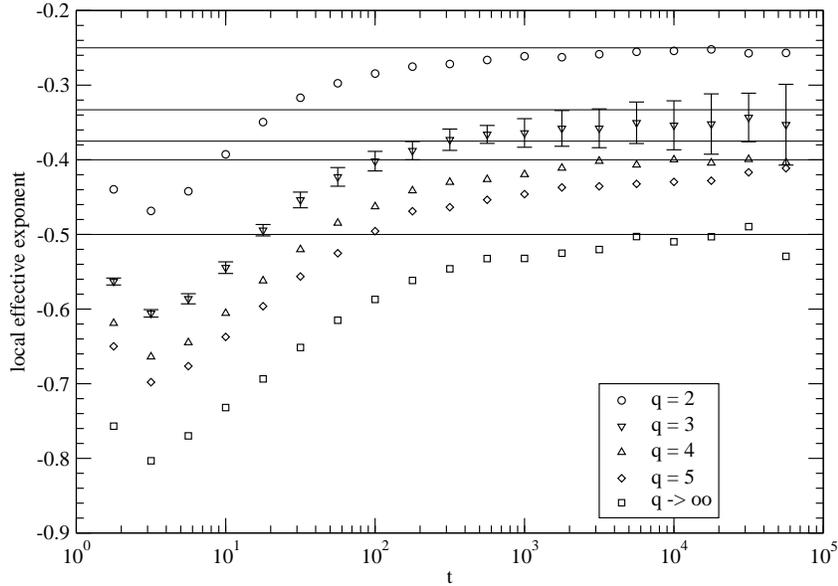}
\caption{\label{3lMAM} Local effective decay exponent $-\alpha(t)$ for the MAM 
on a three-lane system for $q=2$, $3$, $4$, and $5$ (random initial conditions,
equal initial particle numbers for each species). The data were obtained on
$3 \times 100000$ lattice sites, and are averaged over $50$ runs. For the $q=3$
data, we display the statistical error bars. The straight lines correspond to 
the values $\alpha(q,1) =(q-1)/(2q)$ as asymptotically predicted by the 
one-dimensional SMAM.}
\end{center}
\end{figure}

At last, we were interested in the MAM behaviour on $M$ coupled one-dimensional
chains (with periodic boundary conditions along and perpendicular to the 
chains), which we refer to as $M$-lane systems. Within our systematic and 
statistical errors, and in the time window accessible to our simulations, our 
Monte Carlo data for $M=2$ and $M=3$ as depicted in Figs.~\ref{2lMAM} and 
\ref{3lMAM} yield the same results as for a single one-dimensional chain, 
perhaps with somewhat different initial crossover regimes. Although particles 
of different species may now bypass each other, segregation still occurs for 
$q = 2$, $3$, $4$, and $5$ on such few coupled chains, in contrast to the truly
two-dimensional system. 

\section{Conclusion}
\label{secconclusion}

In this paper we have presented the asymptotic long-time behaviour of the 
particle density in the $q$-species pair annihilation model in $d$ dimensions. 
Renormalisation group arguments show that when $q\geq 3$, the total particle 
density in $d>2$ dimensions decays according to a universal power law 
$\rho(t) \sim t^{-\alpha}$, with an exponent $\alpha=1$ that is independent of 
$q$ and $d$. At the upper critical dimension $d_c=2$, this mean-field result
is supplemented by logarithmic corrections, {\it viz.} 
$\rho(t) \sim t^{-1} \ln t$. In dimensions $d\geq 2$, particle segregation 
occurs only for the two-species system ($q=2$), namely for $d<4$, resulting in 
a slower decay with $\alpha(2,d)=d/4$ \cite{ovchinnikov78,toussaint83,lee95}. 
The case $q=2$ turns out to be quite special owing to the presence of a 
{\it local} conservation law for the difference of the two particle densities.
Although consistent with the theoretical predictions, our numerical 
investigations in $d\geq 2$ could naturally be improved by considering larger 
systems, longer simulation times, and perhaps a more sophisticated algorithm as
developed for the one-dimensional case in Ref.~\cite{zhong03}.

The topological constraints in one dimension, however, generically induce 
species segregation, and thereby yields a distinct value for the exponent 
$\alpha$ for each $q$. A simplified, deterministic cellular automaton version 
of the model allows us to find $\alpha(q,1) = (q-1)/(2q)$. This result has been
confirmed numerically through Monte Carlo simulations, which suggests the 
equivalence between the original stochastic model and the simplified version 
at least in the accessible time regimes. We have not succeeded in calculating 
the value of $\alpha(q,1)$ using field theory methods, which remains an open 
but difficult task. 

It would also be interesting to study different versions of the MAM in more
detail \cite{deloubriere02}. For instance, the $q$-species {\it cyclic} 
annihilation model ($q$-CAM) with reactions $A_1+A_2\to\varnothing$, 
$A_2+A_3\to\varnothing$, \ldots, $A_q+A_1\to\varnothing$ represents cyclic 
chemical reactions. For low values of $q$, its properties are easily stated.
First of all, the $q$-CAM and $q$-MAM are straightforwardly equivalent when 
$q=2$ and 3. Then for $q=4$, species 1 and 3 behave in exactly the same manner,
and so do species 2 and 4. Hence their respective distinction is unnecessary,
and the problem is reduced to the 2-MAM, $A+B\to\varnothing$. In the $q=\infty$
limit, however, the CAM becomes a purely diffusing system with no reaction 
occurring at all. It is then for $5\leq q<\infty$ that the $q$-CAM and $q$-MAM 
can be expected to exhibit different behaviour that poses an intriguing problem
for further investigation.

\section*{Acknowledgements}

This research has in part been supported by the National Science Foundation
through grant nos. DMR-0075725 and DMR-0308548, as well as by the Bank of
America Jeffress Memorial Trust, grant no. J-594. 
We thank Paul Krapivsky for bringing this problem to our attention, and
gladly acknowledge helpful discussions with Dani ben-Avraham, John Cardy, 
Manoj Gopalakrishnan, Sid Redner, Jaime Santos, Beate Schmittmann, 
Fr\'ed\'eric van Wijland, Ben Vollmayr-Lee, and Royce Zia.

\section*{Appendix A: Solution of the mean-field rate equations
		      (\ref{rateequations}) and (\ref{mfnonsymmeq})}  

We are interested in the solution of the mean-field rate equations
(\ref{rateequations}) with initial conditions that break the permutation 
symmetry between the $q$ particle species. As shown in the main text, this 
leads to a system of $p>1$ equations (\ref{mfnonsymmeq}) with initial 
conditions $\rho_k(0)=\rho_k^*$ ordered according to 
$0<\rho_1^*<\rho_2^*<\ldots<\rho_p^*$, and where the multiplicities $n_k$
obviously satisfy the constraint $\sum_{k=1}^p n_k=q$. (Notice that the case
$p=2,\,q=2$, and hence $n_1=n_2=1$, 
corresponds to the process $A+B\to \varnothing$ with unequal initial
densities.) We will now obtain the full solution of this system of ordinary 
differential equations. Throughout this appendix, asterisks will refer to 
constants constructed from the initial values. For all $1\leq k\not= n\leq p$ 
we have $\rho_n\not=\rho_k$ and may define 
\begin{equation}
s_{kn}\equiv\frac{\rho_k}{\rho_n-\rho_k} \ ,
\label{defratios}
\end{equation}
and from Eq.~(\ref{mfnonsymmeq}) it follows that these quantities satisfy
\begin{equation}
\frac{\dd s_{kn}(t)}{\dd t}=-\rho_n(t) \, s_{kn}(t) \ .
\label{dsipdt}
\end{equation}
Specifically, consider the $p-1$ variables $s_{kp}$ ($1\leq k\leq p-1$) whose 
initial values $s_{kp}(0)=s_{kp}^*\equiv\rho_k^*/(\rho_p^*-\rho_k^*)$ are 
ordered according to
\begin{equation}
\label{ordersip}
0 < s_{1p}^* < s_{2p}^* < \ldots < s_{p-1,p}^* \, .
\end{equation}
Eq.~(\ref{dsipdt}) shows that the $s_{kp}(t)$ can only decrease with time but 
must remain positive. Upon dividing two of the equations (\ref{dsipdt})
we find $\dd s_{kp}/\dd s_{\ell p}=s_{kp}/s_{\ell p}$, which implies that
\begin{equation}
\frac{s_{kp}}{s_{\ell p}}=\frac{s_{kp}^*}{s_{\ell p}^*}\equiv c_{k\ell p}^* \ ,
\quad 1\leq k,\ell\leq p-1 \, .
\label{constants}
\end{equation}
These ratios, of which $p-2$ are independent, constitute constants of the 
motion of the nonlinear system of ordinary differential equations 
(\ref{mfnonsymmeq}). 

Before continuing we briefly elaborate on these constants of the motion. 
Reasoning as before one easily shows that in fact all ratios 
$c_{k\ell n}\equiv s_{kn}/s_{\ell n}$ (with $k,\ell,n$ all different) are 
constants of the motion; but they may be expressed in terms of the $p-2$ 
already found. There is one more constant of the motion which we do not need
explicitly in our present calculation but mention for completeness. It is
\begin{equation}
C_n=\rho_n^{q-1}\prod_{\ell=1}^p \rho_{\ell}^{-n_{\ell}} \,
\Big(\prod_{\ell=1\atop{\ell\neq n}}^p\,s_{\ell n}\Big)^{(q-1)/(p-1)}
\end{equation}
where $n$ is arbitrary ($C_m$ and $C_n$ with $m\neq n$ may be expressed through
each other with the aid of the $c_{k\ell n}$). Its constancy may be verified by
explicit calculation. $C_n$ differs from the $c_{k\ell n}$ in that it is 
proportional to a nonzero power of the densities. For the fully nondegenerate 
case ($p=q$ and hence all $n_k=1$) our $C_1$ reduces to the constant of the 
motion found in Ref.~\cite{benavraham86}. We emphasise, finally, that when one 
goes beyond the present mean-field approach, these constants of the motion 
{\it do not} correspond to local conservation laws, {\it except} in the case of
$q=p=2$, where $C_1=1/(\rho_1-\rho_2)$.

As a consequence of Eq.~(\ref{constants}), the knowledge of the time evolution 
of the $s_{kp}$ for a single value of $k$, say $s_{1p}(t)$, suffices. After 
dividing Eq.~(\ref{dsipdt}) for $k=1$ and $n=p$ by (\ref{mfnonsymmeq}) for 
$k=p$ we obtain
\begin{equation}
\frac{\dd s_{1p}}{\dd\rho_p}=
\frac{s_{1p}}{\sum_{\ell=1}^{p-1}n_\ell\rho_\ell + (n_p-1)\rho_p} \ .
\label{dsdlogrho}
\end{equation}
From Eqs.~(\ref{defratios}) and (\ref{constants}) we have
\begin{equation}
\rho_\ell=\rho_p \, \frac{s_{1p}}{s_{1p}+c_{1\ell p}^*} \ , \quad 
\ell=1,\ldots,p-1 \, .
\label{rjps1p}
\end{equation}
We next use Eq.~(\ref{rjps1p}) in (\ref{dsdlogrho}) to eliminate 
$\rho_1,\ldots,\rho_{p-1}$ in favour of $\rho_p$. This results in an 
expression for $\dd s_{1p}/\dd\rho_p$ in terms of $s_{1p}$ and $\rho_p$. 
Integration of this expression yields $\rho_p$ as a function of $s_{1p}$. 
Explicitly,
\begin{equation}
\rho_p=\rho_p^*\,F(s_{1p}) \, ,
\label{rhops1p}
\end{equation}
where
\begin{equation}
F(s_{1p})=\left(\frac{s_{1p}}{s_{1p}^*}\right)^{n_p-1}\ \prod_{\ell=1}^{p-1}
\left(\frac{s_{1p}+c_{1\ell p}^*}{s_{1p}^*+c_{1\ell p}^*}\right)^{n_\ell} .
\label{defFsip}
\end{equation}
Combining (\ref{rhops1p}) with Eq.~(\ref{dsipdt}) for $k=1$ and $n=p$ produces 
the time evolution equation for $s_{1p}$,
\begin{equation}
\frac{\dd s_{1p}}{\dd t}=-\rho_p^*\,s_{1p}\,F(s_{1p}) \, .
\label{ds1pdtF}
\end{equation}

The full solution is now to be obtained as follows. Solving the differential
equation (\ref{ds1pdtF}) yields $s_{1p}(t)$, which substituted in 
(\ref{rhops1p}) yields $\rho_p(t)$. These two results together, when used in 
Eq.~(\ref{rjps1p}), produce $\rho_\ell(t)$ for all $\ell=1,\ldots,p-1$. For the
analysis of (\ref{ds1pdtF}) it is useful to know that the $c_{1\ell p}^*$ are 
all positive. We remark parenthetically that they are in fact ordered according
to $1 = c_{11p} > c_{12p} > \ldots > c_{1,p-1,p} > 0$.

We will satisfy ourselves here to extract from Eq.~(\ref{ds1pdtF}) the 
behaviour of $s_{1p}(t)$ for asymptotically large times. In that limit we may 
set $s_{1p}=0$ inside the product, whence Eq.~(\ref{ds1pdtF}) reduces to
\begin{equation}
\frac{\dd s_{1p}}{\dd t}=-\rho_p^*\,A_{n_p-1}^*s_{1p}^{n_p} \, ,
\label{aspteq}
\end{equation}
where one may verify without great effort that
\begin{equation}
A_{n_p-1}^*=(s_{1p}^*)^{-n_p+1}\,
\prod_{\ell=1}^{p-1}(1-\rho_\ell^*/\rho_p^*)^{n_\ell} = (s_{1p}^*)^{-n_p+1}\, 
A^* \, ,
\label{defAn}
\end{equation}
with $A^*$ given by Eq.~(\ref{defA0}). Now two cases have to be 
distinguished:

\noindent {\it Case (i):}\ $n_p=1$.
This is the situation where the highest density is nondegenerate. 
Eq.~(\ref{aspteq}) then leads to
\begin{equation}
s_{1p}(t) \simeq C^* s_{1p}^* \exp(-\rho_p^* A^* t)
\label{sols1paspt}
\end{equation}
where $C^*$ is a constant, and we have factored out $s_{1p}^*$. Combining this 
result with Eq.~(\ref{constants}) we find that in fact
\begin{equation}
s_{kp}(t) \simeq C^* s_{kp}^* \exp(-\rho_p^* A^* t)
\label{solsipaspt}
\end{equation}
for all $k=1,\ldots,p-1$. Upon substituting the asymptotic solution 
(\ref{sols1paspt}) in Eq.~(\ref{rhops1p}) we obtain
\begin{equation}
\rho_p(t)\simeq \rho_p(\infty)=\rho_p^* A^* 
\label{solrhopaspt}
\end{equation}
Further analysis shows that the approach is exponential on a time scale
$\tau=1/\rho_p^*A^*$.
Finally, from the preceding equation and Eqs.~(\ref{defratios}) and 
(\ref{solsipaspt}),
\begin{eqnarray}
\rho_k(t) &\simeq& C^* s_{kp}^* \rho_p^* A^* \exp(-\rho_p^* A^* t) \nonumber\\
&=& C^* \frac{\rho_k^*}{1-\rho_k^*/\rho_p^*}\,A^*\, \exp(-\rho_p^* A^* t) \, , 
\quad 1\leq k\leq p-1 \, .
\label{solrhoiaspt}
\end{eqnarray}
{\it Conclusion:} For $n_p=1$ the nondegenerate densest species $\rho_p(t)$
tends exponentially to a constant $\rho_p^*A^*$ which depends on all initial
densities $\rho_k^*$ according to Eq.~(\ref{defA0}). The densities of the other
species all tend to zero exponentially. The characteristic time scale is 
$\tau=1/(\rho_p^*A_0)$. 

\noindent {\it Case (ii)}:\ $n_p>1$.
In this situation the highest density is $n_p$-fold degenerate. 
Eq.~(\ref{aspteq}) now leads to
\begin{equation}
s_{1p}(t) \simeq  s_{1p}^*[(n_p-1)\rho_p^* A^* t]^{-1/(n_p-1)} \, ,
\label{sols1pasptbis}
\end{equation}
where we have again factored out a $s_{1p}^*$. Combining this with 
Eq.~(\ref{constants}) we find that in fact
\begin{equation}
s_{kp}(t) \simeq s_{kp}^*[(n_p-1)\rho_p^* A^* t]^{-1/(n_p-1)}
\label{solsipasptbis}
\end{equation}
for all $k=1,\ldots,p-1$. Upon substituting the asymptotic solution 
(\ref{sols1pasptbis}) in Eq.~(\ref{rhops1p}) we get
\begin{equation}
\rho_p(t)\simeq \rho_p^*\,(s_{1p}(t)/s_{1p}^*)^{n_p-1} A^*=[(n_p-1)t]^{-1}\, ,
\label{solrhopasptbis}
\end{equation}
which is the mean-field behaviour. 
Finally, from the preceding equation and Eq.~(\ref{solsipasptbis}),
\begin{equation}
\rho_k(t) = \frac{\rho_k^*}{\rho_p^*-\rho_k^*}[\rho_p^* A^*]^{-1/(n_p-1)}\,
[(n_p-1)t]^{-1-1/(n_p-1)} \, , \ 1\leq k\leq p-1 \, .
\label{solrhoiasptbis}
\end{equation}
{\it Conclusion:} For $n_p>1$ the densest species, which are $n_p$-fold
degenerate, tend to zero as $t^{-1}$. The other species tend to zero with a 
faster power, {\it viz.} as $t^{-1-1/(n_p-1)}$.

\section*{Appendix B: Mapping onto an antiferromagnetic spin chain in $d=1$}

Our goal is to construct a field theory in which the specificity of the
one-dimensional case becomes manifest. To this end, we recall that the
single-species annihilation reaction on a lattice can be mapped onto `quantum' 
antiferromagnetic $XXZ$ spin chains, provided we employ site occupation 
restrictions, {\it i.e.}, we allow at most a single particle per lattice site 
$x$ ($n_x = 0$ or $1$) \cite{alcaraz94, henkel97}. For annihilation processes 
that asymptotically approach the empty, absorbing state, this restriction 
should not matter for the long-time behaviour. Moreover, the RG fixed point for
the renormalised reaction rate in $d<2$ dimensions actually corresponds to an 
infinite bare annihilation rate, which precludes multiple site occupancy 
\cite{lee94}. Instead of using bosonic creation and annihilation operators 
acting on a Hilbert space that allows arbitrarily many particles per site, we 
define lowering and raising operators $\sigma^-$ and $\sigma^+$ that operate on
the two possible states on each site,
\begin{equation}
\sigma^-|0\rangle=0 \, , \ \sigma^+|0\rangle=|1\rangle \, , \ 
\sigma^-|1\rangle=|0\rangle \, , \ \sigma^+|1\rangle=0 \, ,
\end{equation}
and obey the standard anticommutation relations for the spin-$1/2$ algebra.

For the $q$-MAM, we naturally introduce $q$ operators $\sigma^\pm_{x,i}$ which
create / destroy a particle of species $i$ on site $x$. It is now a 
straightforward task to rewrite the classical master equation in the form
$\p_t |P(t)\rangle=-\hH |P(t)\rangle$, with a formal state vector $|P(t)\rangle
=\sum_{\{ n_{x,i}= 0,1 \}} P(\{ n_{x,i} \},t)|\{ n_{x,i} \}\rangle$ and the
pseudo-Hamiltonian $\hH = \sum_{x,i}(\hat{h}^{\text{\tiny{diff}}}_{x,i} + 
\sum_{j\neq i}\hat{h}^{\text{\tiny{ann}}}_{x,i,j})$, with the diffusion part
\begin{equation}
\hat{h}^{\text{\tiny{diff}}}_{x,i} = - \frac{D}{2} \left[ 2\hat{n}_{x,i}
(\hat{n}_{x+1,i}-1) + \sigma^+_{x,i}\sigma^-_{x+1,i} + \sigma^-_{x,i}
\sigma^+_{x+1,i} \right] \, ,
\end{equation}
and the annihilation contribution
\begin{equation}
\hat{h}^{\text{\tiny{ann}}}_{x,i,j} = \frac{\lambda}{2} \left[ \hat{n}_{x,i}
\hat{n}_{x+1,j} + \hat{n}_{x,j}\hat{n}_{x+1,i} - \sigma^-_{x,i}\sigma^-_{x+1,j}
- \sigma^-_{x,j}\sigma^-_{x+1,i} \right] \, ,
\end{equation}
where we have defined $\hat{n}_{x,i}=\sigma^+_{x,i}\sigma^-_{x,i}$, and where 
the summations over $i$ and $j$ extend from $1$ to $q$. We remark here that the
exclusion constraint is imposed solely on particles of the same species. Since
the RG fixed point from the `bulk' renormalisation of $\lambda$ in the bosonic
field theory describes an effectively infinite microscopic reaction rate, we 
expect that even particles of different species will asymptotically be 
prevented from occupying identical lattice sites at the same (late) time.

By means of a Jordan-Wigner transformation, one could now proceed to build a 
fermionic field theory from the spin-$1/2$ pseudo-Hamiltonian $\hH$. Indeed, 
the single-species pair annihilation model $A+A\to \varnothing$ has been 
analysed in this framework \cite{schutz95,schutz96,santos96,brunel00}. In order
to take into account the topological effects specific to the one-dimensional 
case, we shall follow here the approach of Refs.~\cite{wiegmann88,fradkin88}, 
as described in the Fradkin's textbook \cite{fradkinbook}. For each spin 
$\vec{S}$, with quantisation axis oriented along $\vec{n}_0$ and eigenstates 
$|S, S_z=\pm \hbar/2 \rangle$, we introduce a vector $\vec{n}$ of modulus $1$ 
that defines the spin coherent state
\begin{equation}
|\vec{n} \rangle = \exp \Big( - \frac{i \, \theta}{\hbar} \, \frac{\vec{n}_0 
\times \vec{n}}{\sin \theta} \cdot \vec{S}\Big) \, |S,S_z=\hbar/2 \rangle \, , 
\quad \text{with} \ \cos\theta = \vec{n}_0 \cdot \vec{n} \, .
\end{equation}
Each vector $\vn$ has three components (spin coordinates) that we label as 
$n^\mu$, $n^\nu$, and $n^\rho$. It can be shown that 
$\langle \vn | \vec{S} | \vn \rangle = \vn \, \hbar/2$, and that for two 
different vectors $\vn_1$ and $\vn_2$,
\begin{equation}
\langle \vn_1 | \vn_2 \rangle = \ee^{i \Phi(\vn_1,\vn_2,\vn_0)/2} \
\sqrt{\frac{1+\vn_1 \cdot \vn_2}{2}} \ ,
\end{equation}
where $\Phi(\vn_1,\vn_2,\vn_0)$ represents the oriented surface area of the 
triangle on the unit sphere whose vertices are given by the endpoints of the 
three vectors $\vn_1$, $\vn_2$, and (fixed) $\vn_0$. 

Through manipulation of these coherent states, the average of an observable can
be written as in Eq.~(\ref{obs}). In discrete time (with the time interval 
$[t_0,t_f]$ divided into $N=(t_f-t_0)/\delta t$ time steps, the limit 
$\delta t \to 0$ to be taken in the end) and on the lattice, the ensuing action
takes the form 
\begin{eqnarray}
{\cal S}[\vn]&=&-\frac{i}{2}\sum_{x=1}^L\sum_{i=1}^q\sum_{k=1}^N \Biggl\{ 
\Phi\Bigl(\vn_{x,i}(t_k),\vn_{x,i}(t_{k-1}),\vn_0\Bigr) \nonumber \\ 
&&- \frac{1}{2} \ \ln \frac{1+\vn_{x,i}(t_k)\cdot \vn_{x,i}(t_{k+1})}{2} +
\delta t\, \langle \vn_{x,i}(t_k)|\hH|\vn_{x,i}(t_k) \rangle\Biggr\} \, , 
\label{mspaction} 
\end{eqnarray}
Here, the first contribution (with $\delta t \to 0$) is of topological
origin, and usually called Wess-Zumino term \cite{fradkinbook}. It is solely 
responsible for the differences between quantum ferromagnets and 
antiferromagnets, as well as for the remarkable qualitative distinction that 
emerges between half-integer and integer antiferromagnetic spin chains. But 
notice its {\it nonlocal} character, since it represents an oriented surface 
area, yielding disjoint sectors each characterised by a certain value of the
corresponding topological charge. This renders a field theory analysis in terms
of a local perturbative expansion futile.

Let us briefly discuss the single-species pair annihilation model 
$A+A\to \varnothing$ (which also describes the $q=\infty$ limit of the 
$q$-MAM) in this framework. One obtains explicitly
\begin{eqnarray}
&&\langle \vn_x(t_k) |\hH| \vn_x(t_k) \rangle = \frac{\lambda-D}{4} \ 
\vn_x(t_k) \cdot \vn_{x+1}(t_k) \\
&&\ +\frac{\lambda}{4}\Bigl[ 2n_x^\rho(t_k)-2n_x^\mu(t_k)\, n_{x+1}^\mu(t_k)
+i\,n_x^\mu(t_k)\, n_{x+1}^\nu(t_k)+i\, n_x^\nu(t_k)\, n_{x+1}^\mu(t_k)\Bigr] 
\, . \nonumber
\end{eqnarray}
In one dimension, the reactions will be diffusion-limited, i.e., the ratio of
the annihilation and diffusion rate $\lambda / D > 1$. Consequently, the 
isotropic Heisenberg spin coupling in the first term is antiferromagnetic (in
contrast to a reaction-limited situation, or to a purely diffusive model)
\cite{schutz95,schutz96,santos96}. Following Ref.~\cite{fradkinbook}, we 
therefore assume that the ground state is represented via the ansatz 
$\vn_x=(-1)^x \vec{m}_x+a_0 \vec{\ell}_x$, where $a_0$ denotes the lattice 
constant. With this change of variables, we may now study the field 
fluctuations around the antiferromagnetic ground state. We now take the 
continuum limit ($a_0 \to 0$ and $\delta t \to 0$), and integrate out the field
$\vec{\ell}$, whereupon after straightforward manipulations we arrive at the
effective action $S=S_{0}+S_{\text{\tiny int}}+S_{\text{\tiny top}}$, with the 
harmonic part
\begin{eqnarray}\fl
\quad S_0=\int\!\dd x \int\!\dd t \Biggl[ \frac{\lambda}{a_0} {m^\mu}^2 - 
\frac{i \lambda}{a_0} m^\mu m^\nu + \frac{i \lambda}{4a_0(\lambda-2D)} \left(
m^\mu\p_t m^\nu - m^\nu \p_t m^\mu \right) \label{actionquad} \\
\fl \qquad -\frac{a_0(\lambda+2D)}{4} \left( \p_x m^\mu \right)^2 + 
\frac{a_0(\lambda-2D)}{4} \left[ \left( \p_x m^\nu \right)^2 + 
\left( \p_x m^\rho \right)^2 \right] + \frac{i \lambda a_0}{2} 
\left( \p_x m^\mu \right) \left( \p_x m^\nu \right) \Biggr] , \nonumber
\end{eqnarray}
and the nonlinear contributions
\begin{eqnarray}\fl
\quad S_{\text{\tiny int}}=\int\!\dd x \int\!\dd t \Biggl[ \frac{1}{16 a_0 
(\lambda-2D)} \left( m^\mu \p_t m^\nu - m^\nu \p_t m^\mu \right)^2 \nonumber \\
\fl \qquad\qquad + \frac{\lambda-2D}{4a_0 (\lambda-4D)^2} \left( m^\nu \p_t 
m^\rho - m^\rho \p_t m^\nu \right)^2 - \frac{D}{2a_0 (\lambda -4D)^2} \left( 
m^\rho \p_t m^\mu - m^\mu \p_t m^\rho \right)^2 \nonumber \\ \qquad\quad
- \frac{i \lambda}{4a_0(\lambda-4D)^2} \left( m^\nu \p_t m^\rho - m^\rho \p_t
m^\nu \right) \left( m^\rho \p_t m^\mu - m^\mu \p_t m^\rho \right) \Biggr] \, .
\label{actionint}
\end{eqnarray}
The topological term reads as in Eq.~(\ref{mspaction}), with $q=1$. We have
furthermore omitted the $\ln$ contribution, since it turns out to be 
irrelevant. The particle kinetics or, equivalently, the spin dynamics can now 
be examined by diagonalising the quadratic part of the action $S_0$. To this 
end, we Fourier-transform Eq.~(\ref{actionquad}) and then determine the 
eigenvalues of the ensuing coupling matrix. Let $k$ and $\omega$ be the Fourier
variables in space and time, respectively. We then find that the modes along 
$m^\rho$ are static. In the $(m^\mu,m^\nu)$ plane, however, there exist 
propagating wave solutions with the long-wavelength and low-energy dispersion
\begin{equation}
q^2\approx\frac{\lambda}{Da_0^2} 
\left( 1 \pm \frac{i \omega}{\lambda-2D} \right) .
\end{equation}
Whereas the particle propagation is purely diffusive for $\lambda=0$, the
presence of the annihilation reactions ($\lambda > 0$) produces an exponential 
decay of the propagator amplitude. Unlike the equivalent bosonic field theory, 
the full action given by Eqs.~(\ref{actionquad}) and (\ref{actionint}) is not 
easily analysed. As in the theory of equilibrium quantum spin chains, other 
methods of analysis such as the Bethe ansatz have proven more fruitful to 
extract the long-time behaviour of such systems.

We now return to the spin representation of the multi-species annihilation 
model. As evident from Eq.~(\ref{mspaction}), each particle species now carries
its own topological charge, described by the corresponding Wess-Zumino term.
Whereas the diffusion part is still diagonal in the species index,
\begin{equation}
\langle \vn_{x,i}(t_k) |\hH| \vn_{x,i}(t_k) \rangle^{\text{\tiny{diff}}} = 
- \frac{D}{4} \vn_{x,i}(t_k) \cdot \vn_{x+1,i}(t_k) \, ,
\end{equation}
the annihilation reactions now couple distinct particle species,
\begin{eqnarray}
&&\langle \vn_{x,i}(t_k) |\hH| \vn_{x,i}(t_k) \rangle^{\text{\tiny{ann}}} = 
\frac{\lambda}{4} \sum_{j\not=i} \Bigl[ \vn_{x,i}(t_k) \cdot \vn_{x+1,j}(t_k) +
n_{x,i}^\rho(t_k)+n_{x,j}^\rho(t_k) \nonumber \\ 
&&\quad -2n_{x,i}^\mu(t_k)\, n_{x+1,j}^\mu(t_k) +i\,n_{x,i}^\mu(t_k)\, 
n_{x+1,j}^\nu(t_k)+i\, n_{x,i}^\nu(t_k)\, n_{x+1,j}^\mu(t_k)\Bigr] \, ,
\end{eqnarray}
and thus in effect link different topological sectors. This fact further 
exacerbates the difficulties with the spin representation of the $q$-MAM, and 
we consider it rather unlikely that this formalism will permit a reliable 
determination of the decay exponent $\alpha(q)$. But the presence of the 
nonlocal topological terms explains why, for integer $q$ at least, the $q$-MAM 
displays quite different behaviour in $d=1$ as compared to higher dimensions. 
For, since we expect the ground state to be characterised by N\'eel 
antiferromagnetic order, already in two dimensions the contributions of 
adjacent chains should cancel the effects of the Wess--Zumino terms at least at
sufficiently long wavelengths \cite{fradkinbook,fnote}. For integer $q$ 
therefore, one would expect special properties, and a dependence of the 
asymptotic scaling regime on the value of $q$, only in one dimension.


\begin{thebibliography}{10}
\bibitem{fnote} 
   Recall, however, that our simulation data for the MAM in two- and three-lane
   systems markedly display the one-dimensional power laws associated with
   species segregation.
\end{thebibliography}
\end{document}